\documentclass[a4paper,11pt]{article}
\pdfoutput=1

\usepackage{jcappub}
\usepackage[T1]{fontenc} 

\usepackage[english]{babel}
\usepackage{natbib,units}
\usepackage{graphicx}
\usepackage{xcolor}
\usepackage{dcolumn}
\usepackage{bm}
\usepackage{amssymb}

\usepackage{verbatim}
\usepackage[utf8]{inputenc}
\usepackage{tikz,hyperref}
\usepackage{dcolumn}
\usepackage{ulem}

\usepackage{amsmath,amssymb}
\usepackage{dsfont}
\usepackage{animate,media9}
\usepackage{slashed}
\usepackage{braket}
\usepackage{subfigure}
\usepackage{slashed}

\usepackage[section]{placeins}

\definecolor{lime}{HTML}{A6CE39}
\DeclareRobustCommand{\orcidicon}{%
        \begin{tikzpicture}
        \draw[lime, fill=lime] (0,0)
        circle [radius=0.16]
        node[white] {{\fontfamily{qag}\selectfont \tiny ID}};
        \draw[white, fill=white] (-0.0625,0.095)
        circle [radius=0.007];
        \end{tikzpicture}
        \hspace{-2mm}
}

\foreach \x in {A, ..., Z}{%
        \expandafter\xdef\csname orcid\x\endcsname{\noexpand\href{https://orcid.org/\csname orcidauthor\x\endcsname}{\noexpand\orcidicon}}
}

\definecolor{lime}{HTML}{A6CE39}
\DeclareRobustCommand{\orcidicon}{%
        \begin{tikzpicture}
        \draw[lime, fill=lime] (0,0)
        circle [radius=0.16]
        node[white] {{\fontfamily{qag}\selectfont \tiny ID}};
        \draw[white, fill=white] (-0.0625,0.095)
        circle [radius=0.007];
        \end{tikzpicture}
        \hspace{-2mm}
}

\foreach \x in {A, ..., Z}{%
        \expandafter\xdef\csname orcid\x\endcsname{\noexpand\href{https://orcid.org/\csname orcidauthor\x\endcsname}{\noexpand\orcidicon}}
}

\title{A Universal Analytic Model for Gravitational Lensing by Self-Interacting Dark Matter Halos}
\author[a,b]{Siyuan Hou$^{\dagger}$\orcidA{}}
\author[a,b]{Daneng Yang$^{\dagger}$\orcidB{}}
\author[c]{Nan Li \orcidC{}}
\author[a,b]{Guoliang Li \orcidD{}}

\footnotetext[1]{$^{\dagger}$These authors contributed equally to this work.}

\emailAdd{syhou@pmo.ac.cn}
\emailAdd{yangdn@pmo.ac.cn}
\emailAdd{nan.li@nao.cas.cn}
\emailAdd{guoliang@pmo.ac.cn}

\affiliation[a]{Purple Mountain Observatory, Chinese Academy of Sciences, Nanjing 210023, China}
\affiliation[b]{School of Astronomy and Space Sciences, University of Science and Technology of China, Hefei 230026, China}
\affiliation[c]{National Astronomical Observatories, Chinese Academy of Sciences, Beijing 100101, China}

\abstract{
We present a model for analytically calculating gravitational lensing by self-interacting dark matter (SIDM) halos. Leveraging the universal behavior of SIDM halos during gravothermal evolution, we calibrate the lensing potential using a fluid simulation, normalizing the evolution time to align with established scenarios. From this potential, we derive explicit equations for the deflection angle and surface density profile, quantifying their deviations from numerical results. Our model builds on the parametric approach of ref.~\cite{yang:2023jwn}, providing refinements in the deep core-collapse regime and enabling more comprehensive lensing studies. 
We explore characteristic lensing features, including critical curves and caustics, for SIDM halos in isolation and within a main halo, tracking their evolution through the gravothermal phase. 
We also examine signatures in the self-similar regime of core collapsed halos and highlight the role of baryonic effects in realistic halos. 
The application of our model extends to generic halos, whose profiles fit one or a superposition of our parametric forms. We make our implementation publicly available on \href{https://github.com/HouSiyuan2001/SIDM_Lensing_Model}{GitHub} to support further research.}

\begin{document}

\maketitle

\tableofcontents 

\section{Introduction}

Gravitational lensing has long been recognized as a crucial tool for probing the matter distribution in the Universe. 
It occurs as a consequence of general relativity, when the path of light from a distant source is bent by the gravitational influence of an intervening massive object, such as a galaxy or a cluster of galaxies~\cite{doi:10.1126/science.84.2188.506,Schneider:2006}. 
This phenomenon provides a direct measurement of the mass distribution along the line of sight, including both visible matter and dark matter, offering a unique observational tool that is largely independent of the detailed dynamics of matter. 
In the standard Cold Dark Matter (CDM) framework, gravitational lensing provides a powerful probe of large- to intermediate-scale structures, with observations aligning well with predictions from cosmological simulations~\cite{Planck:2018lbu,Li:2015xpc,VanWaerbeke:2013eya,1991A&A...248..349B}. 
However, at smaller scales, several anomalies have emerged that challenge the standard CDM paradigm. For instance, the statistics of small-scale lenses from Galaxy-Galaxy Strong Lensing (GGSL) measurements appear to exceed CDM predictions, and the detection of certain dense substructures, known as strong lensing perturbers, suggests densities that are difficult to reconcile with CDM~\cite{Meneghetti:2020yif,Meneghetti:2022apr,Vegetti:2009cz,Vegetti:2012mc,minor:2020hic}. 
Broader observations, such as galactic rotation curves, dynamical mass estimates, and stellar streams, have revealed phenomena that also pose difficulties for the CDM model, suggesting the need to extend or revise our understanding of dark matter's nature~\cite{Bullock:2017xww,Salucci:2018hqu,Gentile:2004tb,vanDokkum:2022zdd,Bonaca:2018fek}. 

Self-Interacting Dark Matter (SIDM) has emerged as a promising framework to address the challenges posed by small-scale structure observations~\cite{Tulin170502358,Adhikari220710638,Spergel9909386}. Through gravothermal evolution, SIDM naturally resolves issues such as the diversity problem in galactic rotation curves~\cite{kochanek:2000pi,kamada:2016euw,ren180805695,santos-santos191109116,correa:2022dey,yang:2022mxl,zavala:2019sjk,kaplinghat:2019svz,turner:2020vlf,slone:2021nqd,silverman:2022bhs}. It also provides potential explanations for a wide range of phenomena, including the properties of satellite galaxies in the Milky Way, dark matter-deficient galaxies, the existence of supermassive black holes at high redshifts, and the mergers of binary supermassive black holes~\cite{nadler:2023nrd,Kong220405981,Zhang:2024ggu,Zhang:2024qem,Zhang:2024qmh,sameie:2019zfo,Yang:2020iya,Kong:2025irr,balberg:2001qg,pollack:2014rja,choquette:2018lvq,feng:2021rst,meshveliani:2022rih,Alonso-Alvarez:2024gdz}. 
Extensive studies, utilizing N-body simulations, conducting-fluid simulations, semi-analytic techniques, and parametric modeling have significantly advanced our understanding of SIDM’s effects, particularly in dark matter-only scenarios~\cite{Despali:2025koj,Correa:2024vgl,mastromarino:2022hwx,Robertson:2017mgj,2017MNRAS.472.2945R,2018ApJ...853..109E,despali:2018zpw,Robertson:2018anx,Zeng:2024xty,Straight:2025udg,Nadler:2024fcs,Han:2023olf,Li:2025rrs,balberg:2002ue,koda11013097,essig:2018pzq,nishikawa:2019lsc,feng:2020kxv,yang220503392,yang:2023jwn,yang:2022zkd,jiang:2022aqw,yang:2023stn,zhong:2023yzk,Gad-Nasr:2023gvf,2018ApJ...853..109E,sameie:2018chj,feng:2020kxv,yang:2023stn,zhong:2023yzk,Fischer:2024dte}.

Importantly, SIDM has shown promise in accounting for strong lensing perturbers and the statistical excess of small-scale lenses.
Ref.~\cite{nadler:2023nrd} conducted a cosmological zoom-in simulation, demonstrating that subhalos under strong SIDM can undergo core collapse, leading to steep surface density slopes that align with values inferred from perturbed lensing images.
For subscale lenses in clusters, ref.~\cite{yang:2021kdf} showed that core-collapsed SIDM halos can enhance the radial GGSL cross section, increasing the number of structures in the inner halo regions.
Ref.~\cite{Dutra:2024qac} further found that the excess in the tangential GGSL cross section can be mitigated if dark matter-only halos undergo core collapse and develop inner density profiles steeper than $r^{-2.5}$.
These studies highlight SIDM as a compelling candidate for enhancing lensing signatures and addressing observational anomalies. 

The study of gravitational lensing relies on understanding mass distribution through key factors such as surface density, lensing potential, and deflection angle. In the context of CDM, simplified parametric profiles---such as the Singular Isothermal Ellipsoid (SIE)~\cite{hinshawGravitationalLensingIsothermal1987, r.kormann*p.schneiderandm.bartelmannIsothermalEllipticalGravitational1993,r.kormann*p.schneiderandm.bartelmannIsothermalEllipticalGravitational1993} and the Navarro-Frenk-White (NFW) profile~\cite{1997ApJ...490..493N,zhaoAnalyticalModelsGalactic1996,ascasibarPhysicalOriginDark2004,dalalOriginDarkMatter2010}---are widely used in lensing studies, allowing for analytic calculations of lensing effects.
In the case of SIDM, although parametric models have been proposed to describe density profiles, predicting lensing effects still requires the numerical evaluation of quantities such as the lensing potential. This reliance on numerical methods poses challenges for modeling lensing in realistic halo populations.

In this work, we propose a new strategy for deriving analytic expressions to model lensing quantities, extending the parametric model introduced in ref.~\cite{yang:2023jwn} to facilitate more comprehensive lensing studies. This model was further developed in ref.~\cite{Yang:2024tba} to incorporate baryons, tested against halos in cosmological simulations in ref.~\cite{Yang:2024uqb}, and integrated into a semi-analytic framework for generating SIDM subhalos in ref.~\cite{Ando:2024kpk}. 
Our method leverages the theoretical approximate universality of gravothermal evolution~\cite{outmezguine:2022bhq,yang220503392,yang:2022zkd,zhong:2023yzk,yang:2023jwn}, which allows us to parametrize an analytic solution for generic SIDM halos. 
We introduce a simple fitting formula to parametrize the lensing potential in a normalized gravothermal phase ($\tau$) and calibrate it against a conducting fluid SIDM simulation.
Our analytic lensing potential enables explicit expressions for the deflection angle and surface density profile, with error estimates across different $\tau$ values. We analyze the evolution of critical curves and caustics throughout SIDM’s gravothermal phases, emphasizing the core-collapse phase, where lensing signatures become more pronounced than in CDM.

This study primarily focuses on dark matter-only SIDM halos. However, we also highlight two major baryonic effects on lensing observations. Our approach is analogous to the method used for the contracted-$\beta$4 profile in ref.~\cite{Yang:2024tba}, which retains the functional form from the dark matter-only case while incorporating the boost in gravothermal evolution and the contraction of the core size.
Although our modeling is preliminary, it underscores the significance of baryons. Further work is needed to develop a more comprehensive understanding.

The paper is organized as follows. 
In section~\ref{sec:data}, we provide the theoretical background and describe the simulation data used for calibration.
Section~\ref{sec:Gravitational Lensing Effects of SIDM Halo} develops an analytic model for SIDM lensing studies and evaluates its performance against numerical results.
Section~\ref{sec:lensingfeatures} explores the lensing signatures derived from our model, focusing on isolated cluster halos and subhalos embedded within central cluster halos. 
Section~\ref{sec:selfsimilar} presents a detailed analysis of lensing signatures of core collapsed halos in the self-similar regime. In section~\ref{sec:baryons}, we explore the impact of baryons, using a contracted density profile as an estimate.
Section~\ref{sec:Conclusions} summarizes our results, discussing potential applications of our model and directions for future development. 
For numerical illustrations, we assume a flat $\Lambda$CDM cosmology with $h = 0.7$, $\Omega_m = 0.3$, $\Omega_\Lambda = 0.7$, and $w_0 = -1$. Our calibrated models are independent of these cosmological parameters. 

\section{Universality and simulation data for model calibration}
\label{sec:data}

\begin{figure}
    \centering
    \includegraphics[width=\linewidth]{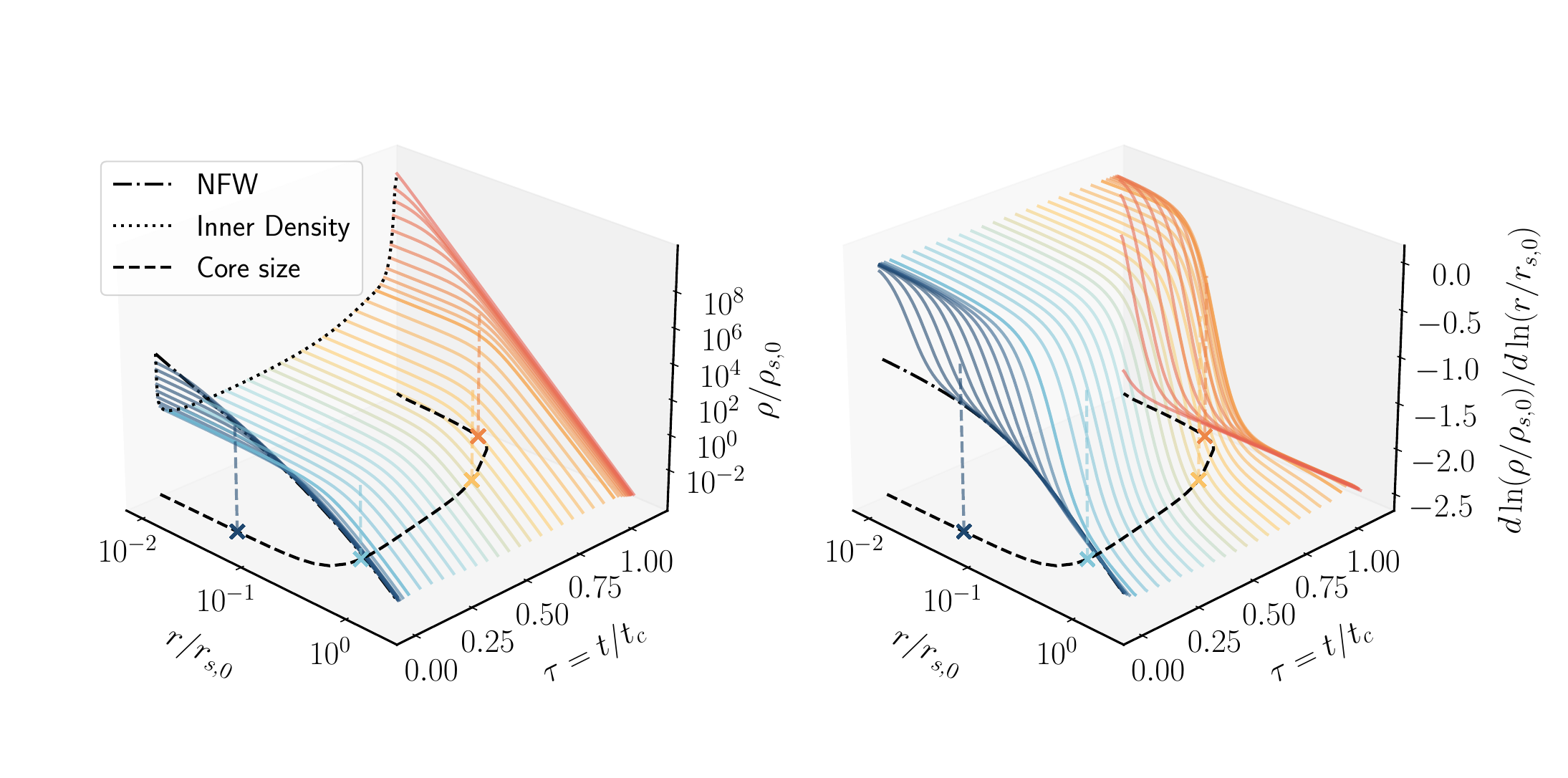}
    \caption{ Visual representation of the universal gravothermal evolution described by the parametric logarithmic density profile (left) and its radial derivatives (right). Assuming spherically symmetric halos, density and radius are normalized by the NFW parameters $\rho_s$ and $r_s$, respectively. Colder (warmer) colors correspond to smaller (larger) $\tau$ values, where $\tau \equiv t/t_c$. The dashed black curves on the bottom planes depict the evolution of the core size $r_c$ in the parametric model, with four specific points at $\tau = 0.001$, $0.15$, $0.8$, and $1.05$ highlighted (crossed) to show their corresponding densities (left) or density slopes (right). A dotted black curve on the left panel highlights the evolution of inner densities at $r = 10^{-2} r_{s,0}$.
    }
    \label{fig:3d-density}
\end{figure}

Gravothermal evolution in the long-mean-free-path (LMFP) regime exhibits a universal behavior when the initial density profiles and evolution times of different halos are appropriately normalized~\cite{Lynden-Bell:1968eqn,balberg:2001qg,outmezguine:2022bhq,yang220503392,yang:2022zkd,zhong:2023yzk,yang:2023jwn}.
This universal behavior has been observed in both conducting fluid simulations and N-body simulations. However, the degree of universality reported in the literature varies: different numerical implementations yield divergent core collapse times, with discrepancies ranging from 10\% to 30\%~\cite{Yang:2024uqb,outmezguine:2022bhq,yang220503392,yang:2022zkd,zhong:2023yzk,yang:2023jwn}.

Studies have investigated the uncertainties in the normalization factor for core collapse time, attributing them to both numerical and physical sources~\cite{Yang:2024uqb,yang:2022zkd,zhong:2023yzk,Fischer:2024eaz,Palubski:2024ibb}. In the fluid approach, the treatment of heat conduction depends on simplified models of heat conduction, including transitions between the short- and long-mean-free-path regimes. 
In the N-body approach, factors such as timestep criteria and the treatment of multiple scatterings introduce uncertainties and energy conservation violations, which undermine universality.

Despite these challenges, several studies have shown that the shape of gravothermal evolution remains largely universal. A parametric model constructed based on this universality has been demonstrated to be effective for both isolated halos and subhalos. Ref.~\cite{Yang:2024tba} offered an intuitive way to understand the reliability of universal behavior: as long as the SIDM effects in a halo scale with the number of scatterings (as in the LMFP regime), these effects can be fully absorbed into the evolution time. This allows for the introduction of a dimensionless gravothermal phase, $\tau=t/t_c$, where $t$ is the halo's evolution time and $t_c$ is the core collapse time.
For dark matter-only halos, the core collapse time can be expressed as~\cite{balberg:2002ue,pollack:2014rja,essig:2018pzq}:
\begin{eqnarray}
\label{eq:tc}
t_{\rm c}  &=& \frac{150}{C} \frac{1}{(\sigma/m) \rho_s r_s} \frac{1}{\sqrt{4\pi G \rho_s}},
\end{eqnarray}
where $\rho_s$ and $r_s$ are the scale parameters of the NFW profile, and $C$ is the normalization parameter. In the parametric model proposed in ref.~\cite{yang:2023jwn}, $C=0.75$ was chosen to match N-body simulations~\cite{koda11013097,essig:2018pzq,nishikawa:2019lsc,yang:2022zkd}.
The concept of universality has been developed in recent years; for more details, see refs.~\cite{yang:2023jwn,zhong:2023yzk,outmezguine:2022bhq,yang220503392,yang:2022zkd}.

We present a visual representation of the universal gravothermal evolution in figure~\ref{fig:3d-density}, which can be interpreted as a general solution for a gravothermal system: Given a halo and an SIDM model, the evolution can be derived by specifying the corresponding $\rho_{s,0}$, $r_{s,0}$, and $t_c$. 
The density profile is normalized by the NFW scale parameters at $\tau=0$, with its evolution scaled by $t_c$. The corresponding logarithmic density slope, $d\ln(\rho/\rho_{s,0})/d\ln(r/r_{s,0})$, is shown in the right panel. At $\tau=0$, the density profile follows the NFW form, characterized by an inner logarithmic slope of $-1$. By $\tau \approx 1.1$, the density transitions to a power-law profile, $r^{-2.2}$, featuring a core-collapsed halo in the LMFP regime~\cite{Lynden-Bell:1980xip,balberg:2001qg}. 
Dotted and dashed lines are used to track the evolution of the inner halo density, $\rho_c$, at $r/r_{s,0} = 0.01$, and the core size, $r_c$, within the parametric model. Following a sharp decline at $\tau \gtrsim 0$, $\rho_c$ reaches its minimum near $\tau \lesssim 0.2$, after which it stabilizes into a quasi-stable core that contracts slowly. As $\tau$ approaches unity, core contraction accelerates significantly, indicating the onset of runaway collapse.

Core-collapsing halos play a critical role in lensing studies, as their denser dark matter distributions are more likely to exceed the critical density for lensing, thereby enhancing lensing effects. However, the parametric model in ref.~\cite{yang:2023jwn} is calibrated using an isolated halo in a high-resolution N-body simulation. As the halo enters the deep core-collapse phase and becomes increasingly dense, the finite softening length in the N-body simulation approaches the scale of the core, making it progressively challenging to accurately model the steep increase in inner density.

\begin{figure}
    \centering
    \includegraphics[width=\linewidth]{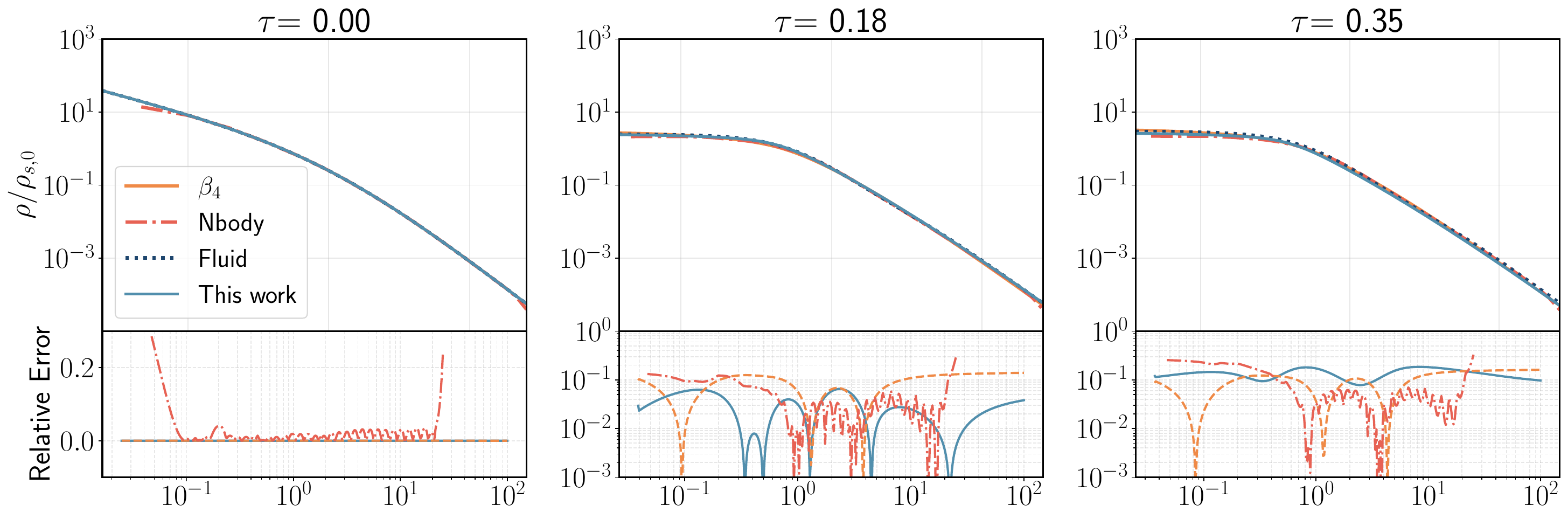}
    \includegraphics[width=\linewidth]{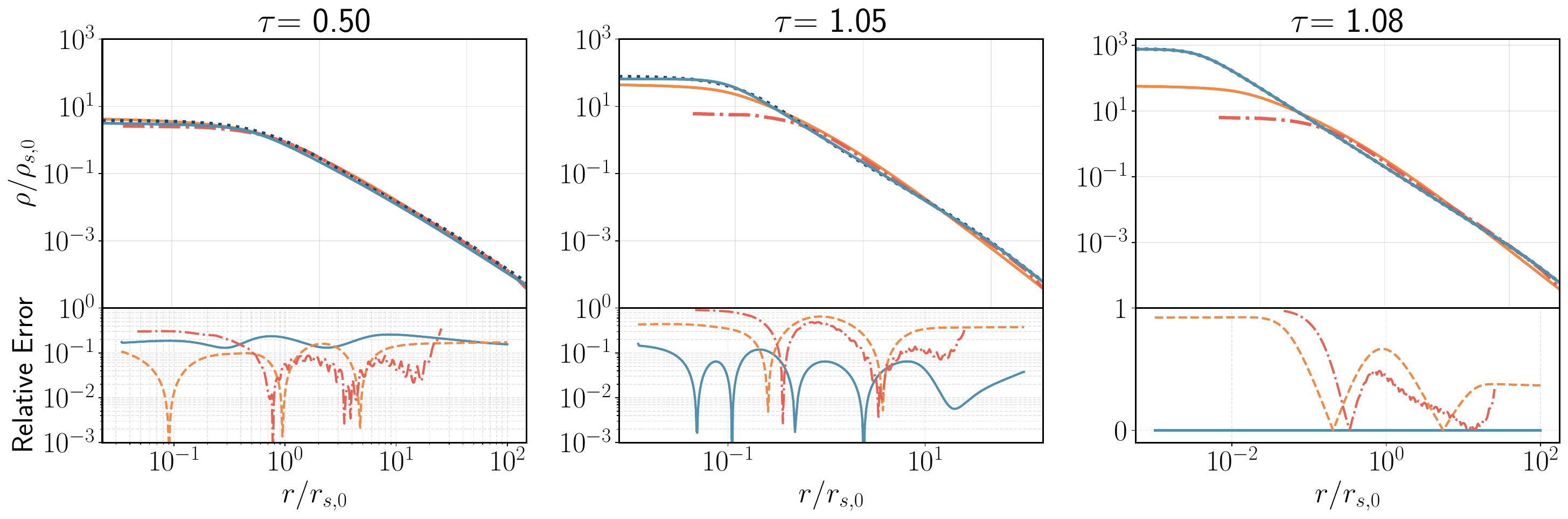}
     \caption{Comparison of SIDM halo density profiles from simulations and model predictions at representative $\tau$ values. The N-body simulation results from ref.~\cite{yang:2023jwn} were used to calibrate the $\beta4$ model (orange), while the fluid simulation based on the program of refs.~\cite{Gad-Nasr:2023gvf,outmezguine:2022bhq} was used to calibrate the PSIDM model (blue) in this work. As discussed in the main text, differences between the previous and new models are minimized for $\tau < 0.7$, beyond which the fluid simulation results are expected to provide greater accuracy, particularly in the inner regions. The lower panels show the relative errors of each model with respect to the fluid simulation results.
     }
     \label{fig:compfluidNbody}
\end{figure}

To enhance the accuracy of modeling in the deep core-collapse phase, we recalibrate a parametric model using high-precision data from fluid SIDM simulations (hereafter referred to as fluid simulations).
We simulate an isolated halo with an initial NFW profile characterized by $\rho_{s,0} = 1.94 \times 10^7~\rm M_{\odot}/kpc^3$ and $r_{s,0} = 10~\rm kpc$, using the fluid simulation program from refs.~\cite{Gad-Nasr:2023gvf,outmezguine:2022bhq}. Assuming universality, we normalize both the simulated halo density profile and its evolution prior to calibration. To address uncertainties in the normalization of $t_c$, we align the gravothermal phase evolution by matching the density profiles with the predictions of the prior parametric model at $\tau \lesssim 0.7$. This alignment is achieved by adopting $C = 0.8$ in the fluid simulation. For practical applications of the calibrated parametric model, one can still use $C = 0.75$ for calculating $t_c$. 
This approach ensures a more accurate estimation of the overall collapse time, as derived from N-body simulations, while leveraging the fluid simulation's capability to accurately model the inner density structure and behavior at large $\tau$ values. By combining the strengths of both methods, this dual calibration framework enhances the reliability of obtained results.

In figure~\ref{fig:compfluidNbody}, we compare the normalized density profiles from the fluid and N-body simulations at representative $\tau$ values. The model predictions from the parametric model of ref.~\cite{yang:2023jwn} ($\beta4$) and our updated model, as described in section~\ref{sec:density}, are also included for reference.
Notably, the two simulations show significant deviations at $\tau > 1$, while they align closely at $\tau < 0.5$. The $\beta4$ model begins to diverge from the N-body simulation at $\tau > 1$, whereas our new model provides an accurate fit to the density profiles across all times.

\section{Constructing an analytic model for gravitational lensing}
\label{sec:Gravitational Lensing Effects of SIDM Halo}

Obtaining lensing predictions from an analytic density profile requires integrating the profile to compute the surface density and solving for the lensing potential. However, analytic solutions for these quantities exist only for certain simple profiles. To preserve the high fidelity of modeling the profiles while simultaneously enabling analytic expressions for lensing-specific quantities, we propose a novel strategy grounded in the universality of gravothermal evolution.

We begin by introducing a simple analytic model for the lensing potential. To minimize uncertainties in modeling lensing effects, we compute the normalized deflection angle, $\hat{\alpha}$, analytically from this lensing potential. We then determine the evolution of the parameters in the lensing potential model by fitting $\hat{\alpha}$ to numerically obtained results. The simplicity of our analytic lensing potential also allows us to derive the surface density in closed form.

By calibrating the model based on the deflection angle, we simultaneously achieve predictions for the lensing potential and surface density. These predictions are subsequently validated against numerically obtained results to ensure their accuracy and reliability.

\subsection{The lensing model}

In lensing studies, the surface mass density of a lens must exceed a critical density,
$$ 
\Sigma_{cr} = \frac{c^2}{4\pi G}\frac{D_S}{D_L D_{LS}}
$$
to produce significant lensing phenomena such as multiple images, arcs, or Einstein rings. Here, $D_S$, $D_{LS}$ are the angular diameter distances to the source and between the lens and the source, respectively. 
Given the surface mass density of the lens, the lensing potential, $\Psi(R)$, can be determined by solving the Poisson equation,
$$
\nabla^2\Psi(R) = 2\frac{\Sigma(R)}{\Sigma_{\mathrm{cr}}}.
$$
For a circularly symmetric system, the lensing potential can be expressed as:
\begin{equation}
\Psi(R) = \frac{2}{\Sigma_{\mathrm{cr}}} \int_0^R s \Sigma(s) \ln \left(\frac{R}{s}\right) ds,
\end{equation}
where both $\Psi$ and $\Sigma$ are functions of the two-dimensional radius $R$. 

To construct a universal model, we define the following dimensionless quantities:
\begin{eqnarray}
\hat{\Psi}\equiv \frac{\Psi \Sigma_{cr}}{\rho_{s,0} r_{s,0}^3}, \text{       } \hat{\Sigma} \equiv \frac{\Sigma}{\rho_{s,0} r_{s,0}^2}, \\ \nonumber
\end{eqnarray}
where $\hat{\Psi}$ and $\hat{\Sigma}$ are expressed as functions of the normalized two-dimensional radius, $\hat{R} = R/r_{s,0}$. 
Using these dimensionless variables, the lensing potential becomes
\begin{gather}
\hat{\Psi}(\hat{R}) = 2 \int_0^{\hat{R}} \hat{s} \hat{\Sigma}(\hat{s}) \ln \left(\frac{\hat{R}}{\hat{s}}\right) d \hat{s}.
\end{gather}

Using fluid simulation data for the density profiles, we numerically compute the corresponding lensing-specific quantities. To parameterize the lensing potential, we propose the following functional form:
\begin{equation}
    \hat{\Psi}(\hat{R}) = a \ln (1 + b \hat{R} + c \hat{R}^2)^c - \ln (p \hat{R} + 1)^s, \label{eq:SIDM_lens_potential}
\end{equation}
where $a$, $b$, $c$, $p$, and $s$ are functions of $\tau$.

To minimize errors in modeling the lensing effect, which is determined by the deflection angle, we extract the evolution of the parameters $a$, $b$, $c$, $p$, and $s$ by fitting to the normalized deflection angle $\hat{\alpha} = \partial \hat{\Psi}/\partial \hat{R}$. 
We use the fitted results from a previous snapshot as the initial guess for the subsequent fit, ensuring a smooth evolution of the fitting parameters. To capture the evolution of these trajectories, we introduce fitting functions and obtain the following results:
    \begin{equation}
        \label{eq:SIDM_lens_potential_parameters}
        \begin{aligned}
        a(\tau) &= 1.113 - 1.262 \cdot \tau^{0.1} + 2.876 \cdot \tau^{0.2} - 3.616 \cdot \tau^{0.5} 
        + 2.778 \cdot \tau^{0.9} - 0.4159 \cdot \tau^2 \\
        &\quad + 0.01357 \cdot \tau^{12} - 0.03862 \cdot \tau^{20} - 0.0002633 \cdot \tau^{87}, \\
        b(\tau) &= 6.597 + 2.213 \cdot \tau^{0.01} - 9.563 \cdot \tau^{0.2} + 9.104 \cdot \tau^{0.6} 
        - 4.013 \cdot \tau^{0.9} + 3.268 \cdot \tau^2 + 2.076 \cdot \tau^{12} \\
        &\quad + 0.1184 \cdot \tau^{51} - 0.0002105 \cdot \tau^{109} + 4.214 \cdot 10^{-7} \cdot \tau^{203}, \\
        c(\tau) &= 1.793 + 0.5108 \cdot \tau^{0.1} - 1.430 \cdot \tau^{0.2} + 1.392 \cdot \tau^{0.4} 
        - 0.7858 \cdot \tau^{0.9} + 0.1604 \cdot \tau^2 \\ &\quad + 0.03146 \cdot \tau^{12} 
         + 0.0007148 \cdot \tau^{67}, \\
        p(\tau) &= 6.943 + 4.967 \cdot \tau^{0.2} - 12.53 \cdot \tau^{0.3} + 7.546 \cdot \tau^{0.8} 
        + 3.212 \cdot \tau^2 + 2.863 \cdot \tau^{11} - 0.3265 \cdot \tau^{22} \\
        &\quad + 0.3093 \cdot \tau^{37}, \\
        s(\tau) &= 1.825 - 0.5991 \cdot \tau^{0.2} + 0.8449 \cdot \tau^{0.3} - 0.4346 \cdot \tau^{0.8} 
        + 0.08176 \cdot \tau^2 + 0.01675 \cdot \tau^{11} \\ &\quad - 0.001751 \cdot \tau^{22} 
        + 0.003828 \cdot \tau^{37}.
        \end{aligned}
    \end{equation}

\begin{figure}
    \centering
    \includegraphics[width=\linewidth]{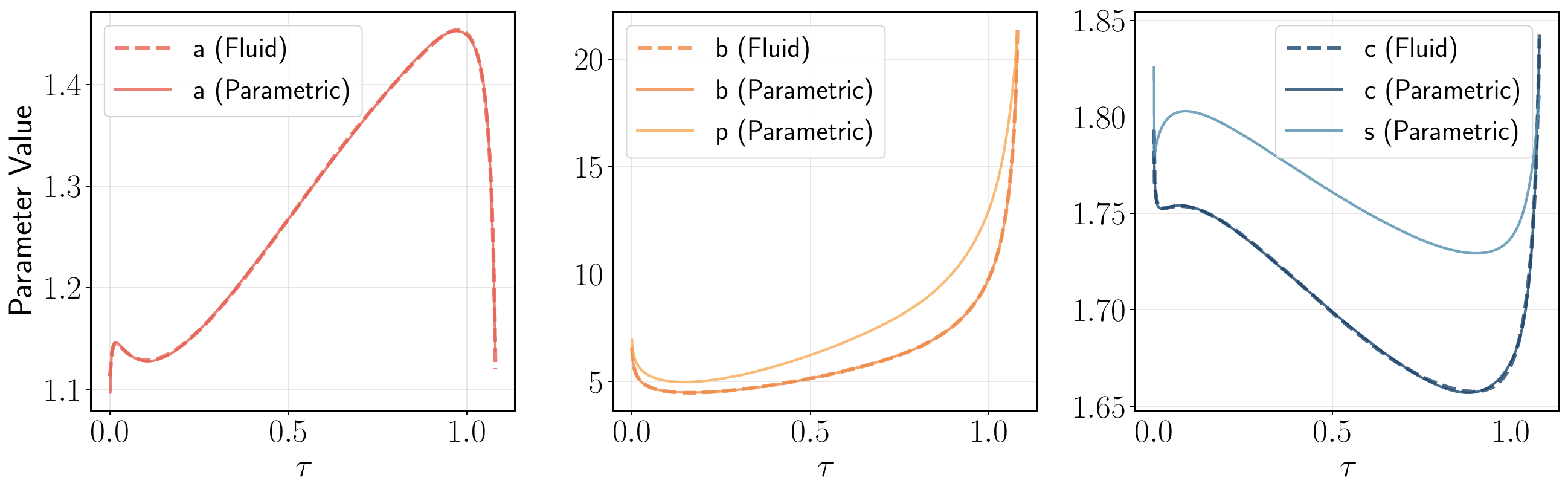}
    \caption{The lensing model parameters $a$, $b$, $c$, $p$, and $s$ as functions of $\tau$, depicted for the fluid simulation (dashed lines) and the calibrated parametric model (solid lines). Note that $p(\tau)$ and $s(\tau)$ have no counterparts in the fluid simulation, as they were fixed during a preliminary calibration to better control the fitting of the other parameters. In all cases, the relative differences of the parametric model to the fluid results are consistently below $10^{-4}$. }
     \label{fig:abc}
\end{figure}
    
Figure~\ref{fig:abc} compares the fitted curves, as defined by the equations, with those obtained from the deflection angle fitting. The three panels showcase the remarkable accuracy in fitting the $a$ (left), $b$ (middle), and $c$ (right) parameters. 
The $p$ and $s$ parameters do not have counterparts from the fluid simulations. This is because their evolution was fixed based on a rough pre-simulation to better control the fitting of the $a$, $b$, and $c$ parameters. Nevertheless, the middle and right panels display the evolution of the $p$ and $s$ parameters in lighter colors, respectively.

Based on these evolutionary trajectories, the evolution of the normalized deflection angle and surface density can also be determined. The normalized deflection angle is given by:
\begin{equation}
    \label{eq:SIDM_deflection_angle}
    \hat{\alpha}(\hat{R})= \frac{\partial \hat{\Psi}(\hat{R}) }{\partial \hat{R}} = \frac{a\cdot c \left(b + 2 c \hat{R}\right) \ln^{c-1}\left(b \hat{R} + c \hat{R}^2 + 1\right)}{b \hat{R} + c \hat{R}^2 + 1} - \frac{p\cdot s \ln^{s-1}\left(p \hat{R} + 1\right)}{p \hat{R} + 1}
\end{equation}

The normalized surface density is expressed as: 

\begin{align}
\label{eq:surfacedensity}
    \hat{\Sigma}(\hat{R}) &= \frac{1}{2\hat{R}} \frac{\partial}{\partial\hat{R}} \Bigg[\hat{R} \cdot \hat{\alpha}(\hat{R})\Bigg] \notag \\
    &= \frac{1}{2} \Bigg[
    \frac{2 a c^2 \ln^{c-1}(D_1)}{D_1}
    + \frac{a (c - 1) c (b + 2 c \hat{R})^2 \ln^{c-2}(D_1)}{D_1^2} \notag \\
    &\quad - \frac{a c (b + 2 c \hat{R})^2 \ln^{c-1}(D_1)}{D_1^2}
    + \frac{p^2 s \ln^{s-1}(D_2)}{D_2^2}
    - \frac{p^2 (s - 1) s \ln^{s-2}(D_2)}{D_2^2}
    \Bigg] \notag \\
    &\quad + \frac{1}{2 \hat{R}} \Bigg[
    \frac{a c (b + 2 c \hat{R}) \ln^{c-1}(D_1)}{D_1}
    - \frac{p s \ln^{s-1}(D_2)}{D_2}
    \Bigg],
\end{align}
where the following functions have been introduced to simplify the expression:
\begin{align}
    D_1 &= b \hat{R} + c \hat{R}^2 + 1, \quad D_2 = p \hat{R} + 1.
\end{align}

\begin{figure}
    \centering
    \includegraphics[width=\linewidth]{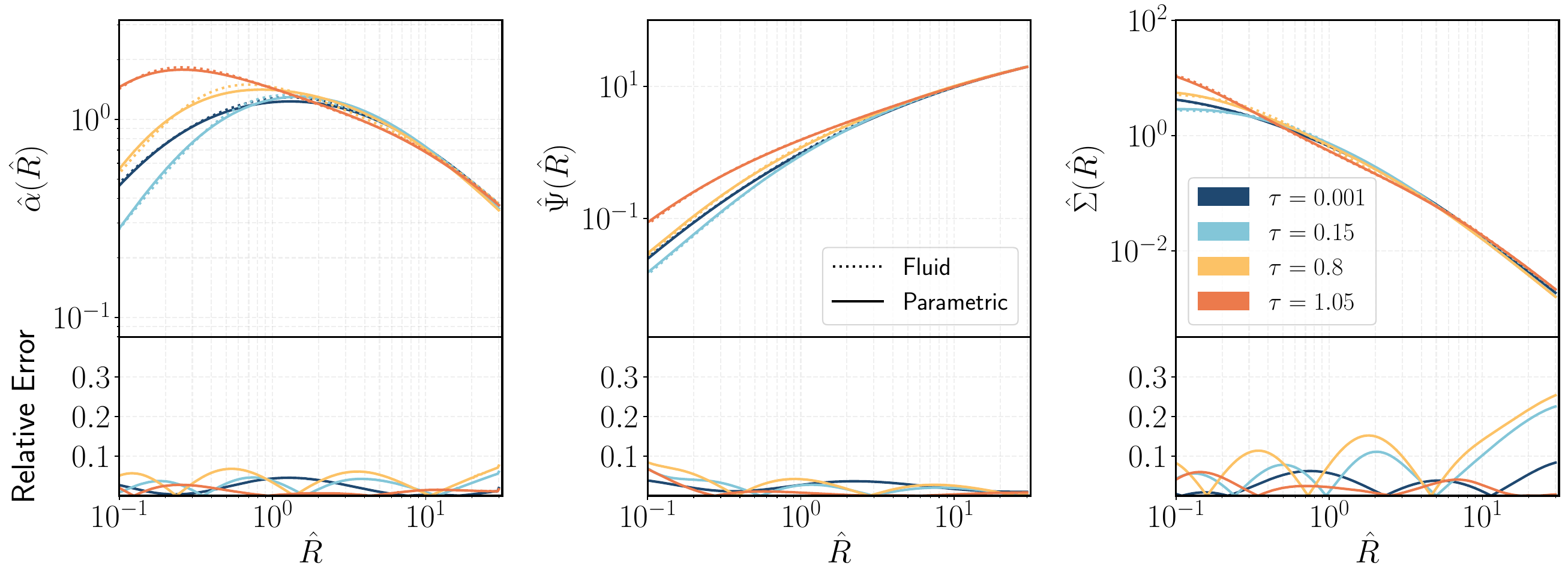}
    \caption{Comparison of simulated (dotted) and model-predicted (solid) normalized deflection angles (left), lensing potential (middle), and surface density (right) at $\tau = 0.001$, $0.15$, $0.8$, and $1.05$. The lower panels show the absolute relative differences between the dotted and solid curves, calculated as ${\rm (Parametric-Fluid)/Fluid}$, highlighting the level of agreement between the simulations and the model predictions.}
    \label{fig:deflection_lensing_density}
\end{figure}

In Fig.~\ref{fig:deflection_lensing_density}, we evaluate the quality of our lensing model by comparing its predictions (solid) with numerical results (dotted) at representative $\tau$ values. 
The three panels, from left to right, display the normalized deflection angles, lensing potentials, and surface densities, respectively. The lower panels of each figure show the absolute relative errors, which remain below 10\% for $\hat{\alpha}$, $\hat{\Psi}$, and mostly below 20\% for $\hat{\Sigma}$ in almost the entire illustrated range of $\hat{R}$. 

\subsection{The density profile model}
\label{sec:density}

Although analytic, the equation for the surface density (Eq.~\ref{eq:surfacedensity}) is already quite complex. Instead of attempting to derive an analytic density profile from the inversion of Eq.~\ref{eq:surfacedensity}, we propose an alternative density profile model.  
This model introduces a core in a manner analogous to refs.~\cite{yang:2023jwn,robertson:2016qef}, but includes two additional parameters, which significantly enhance its ability to model density profiles in the deep core-collapse regime ($\tau\gtrsim 0.7$). The new density profile, which we refer to as the PSIDM profile (short for Parametric SIDM), is given by:
\begin{equation}
    \label{eq:SIDM_density}
    \rho_{\rm PSIDM}(r) = \frac{\rho_s}{\left[\left(\frac{r}{r_s}\right)^4 + \left(\frac{r_c}{r_s}\right)^4\right]^{\frac{\gamma}{4}} \cdot \left[1 + \left(\frac{r}{r_s}\right)^\beta\right]^{\frac{3-\gamma}{\beta}}},
\end{equation}
where $\rho_s$, $r_s$, $r_c$, and the newly introduced parameters $\beta$ and $\gamma$ are all functions of $\tau$. We calibrate the model in a manner similar to the lensing model, obtaining the following evolution trajectories for these parameters:
\begin{small}
\begin{equation}
    \begin{aligned}
    \frac{\rho_s}{\rho_{s,0}} &= \exp\left( 
    7.221 \cdot \tau^{0.2} - 15.49 \cdot \tau^{0.3} + 17.78 \cdot \tau^{0.7} 
    - 721.8 \cdot \tau^{0.8} + 564.0 \cdot \tau - 0.4527 \cdot \tau^6 - 1.242 \cdot \tau^{12} 
    \right), \\
    \frac{r_s}{r_{s,0}} &= 1 + 2.797 \cdot \tau^{0.6} 
    - 2.687 \cdot \tau + 3.421 \cdot \tau^3 + 0.06970 \cdot \tau^{10} + 2.027 \cdot \tau^{17} + 0.03141 \cdot \tau^{80}- 0.02225 \cdot x^{83}, \\
    \frac{r_c}{r_{s,0}} &= 6.046 \cdot \tau^{0.6}-6.957 \cdot \tau^{0.8}  
    + 1.486 \cdot \tau^2 - 0.6234 \cdot \tau^5 + 0.2096 \cdot \tau^{12} 
    - 0.0338 \cdot \tau^{27} + 5.026 \cdot 10^{-5} \cdot \tau^{80}, \\
    \beta &= 1 - 0.4171 \cdot \tau^{0.6} + 0.2024 \cdot \tau^2 + 0.9345 \cdot \tau^{13} - 2.110 \cdot \tau^{26} + 2.008 \cdot \tau^{39} - 0.5762 \cdot \tau^{66} + 0.2962 \cdot \tau^{71}\\
    \gamma &= 2.547 - 2.756 \cdot \tau + 6.118 \cdot \tau^2 - 7.289 \cdot \tau^3 
    + 3.419 \cdot \tau^4 + \frac{1}{\ln(0.001)} (-1.547) \ln(\tau^{1.330} + 0.001) \\
    &\quad + 0.1 \cdot \left( \arctan(3(\tau - 0.5)) - \arctan(-1.5) \right).
    \end{aligned}
\end{equation}
\end{small}

In this model, the parameter $\gamma$ controls the inner density slope when $r_c$ is much smaller than $r_s$. In the outer region, the profile declines as $r^{-3}$, similar to the NFW profile, while the parameter $\beta$ allows for control over the shape in the intermediate regions.

In Fig.~\ref{fig:compfluidNbody}, we present the performance of this model in fitting the fluid simulation results at representative $\tau$ values. We achieve excellent fitting quality across all gravothermal phases in the range $0<\tau<1.08$. However, for $\tau\gtrsim 1.08$, the halo undergoes a runaway collapse, and the density profile approaches a power-law form in the LMFP regime. This abrupt transition is physical and cannot be smoothly captured by fitting functions. In Section 4, we conduct a detailed study of core-collapsed halos in this regime, where we demonstrate that the PSIDM profile still provides an excellent fit.

\section{Lensing features from analytic and numerical evaluations}
\label{sec:lensingfeatures}

We have quantified the differences between our model predictions in the lensing-specific profiles and density profiles with the ones from the fluid simulation data. 
Here, we extend our analysis to evaluate the performance of our model in computing critical curves and caustics, which are essential tools for interpreting observed lensing phenomena and deriving constraints on lens properties.
Critical curves are defined on the lens plane as the loci where the determinant of the Jacobian matrix of the lens equation vanishes, i.e.,
$$
\det\left(\frac{\partial \boldsymbol{\beta}}{\partial \boldsymbol{\theta}}\right) = 0,
$$
where $\boldsymbol{\theta}$ represents the angular position in the lens plane and $\boldsymbol{\beta}$ denotes the corresponding angular position in the source plane. The mapping of critical curves onto the source plane defines caustics, which are curves where lensing magnification formally diverges. Sources positioned near or on caustics exhibit significant magnification and are often multiply imaged.

The numerical results used to validate our models are based on density profiles derived from the fluid simulation. The surface density is computed by integrating the density profile along the line of sight. The deflection angle is determined by calculating derivatives of the lensing potential, which involves solving the Poisson equation. To achieve accurate lensing potential values, we apply a discrete Fast Fourier Transform (FFT) on a $4000 \times 4000$ grid with a side length of $2.5r_{s,0}$, and then focus on the region around the critical curve to refine the analysis. To ensure the accuracy of the numerical results, we apply zero-padding around the edges of the grid by adding a layer of zeros equal to the side length. This approach ensures precise numerical results for the critical curves and caustics. 

\subsection{Isolated halos}

To demonstrate the accuracy of our model when applied to an isolated system, we consider a cluster-scale halo characterized by $\rho_{s,0}=8.92\times 10^{5} ~\rm h^2 \rm M_{\odot}/kpc^3$ and $r_{s,0} = 933~\rm kpc/h$ at various gravothermal phases. In reality, a cluster-scale halo cannot evolve into the core-collapsing phase without violating observational constraints. However, this setup serves as an idealized framework for systematically testing the performance of our model under representative gravothermal phases. 

\begin{figure}
    \centering
    \includegraphics[width=\linewidth]{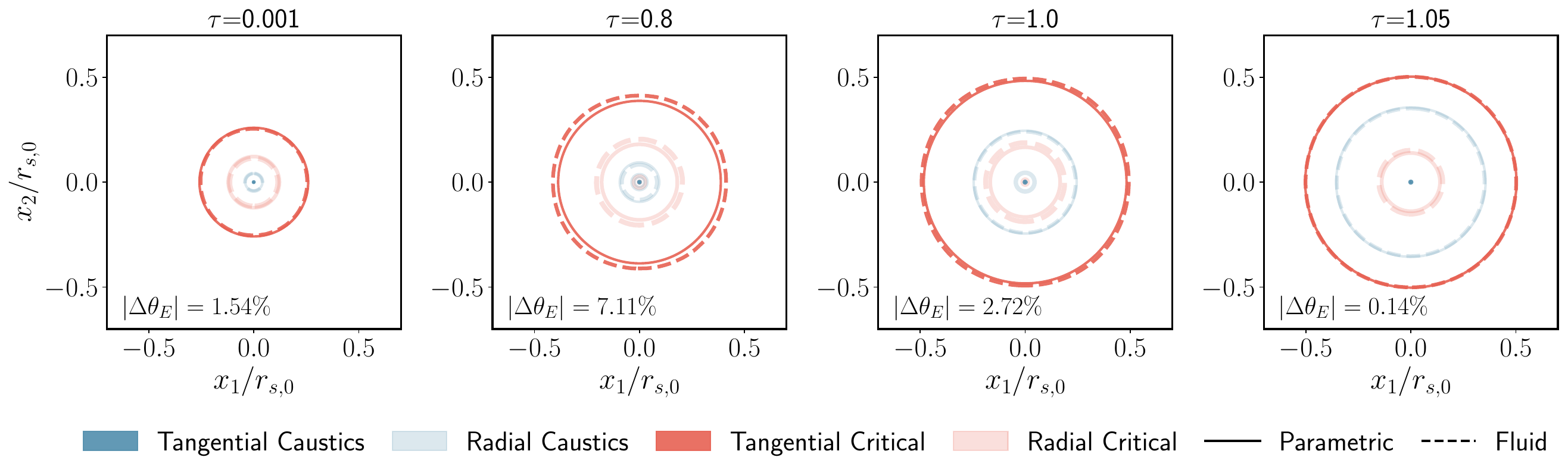}
    \caption{ Critical curves and caustics for a spherical, isolated SIDM halo. From left to right, the results correspond to progressively larger gravothermal phases with $\tau = 0.001$, $0.8$, $1.0$, and $1.05$. Solid curves denote predictions from the parametric model, while dashed curves represent numerical results from the fluid simulation. Both critical curves (red) and caustics (blue) are displayed within the same panels, keeping in mind that they lie on different planes: the lens and the source. Tangential critical curves and caustics are highlighted with darker lines, while radial critical curves and caustics are shown with lighter lines for clarity. $|\Delta \theta_E|$ denotes the absolute difference between the effective Einstein radii obtained from the parametric model and the fluid simulation. 
    }
    \label{fig:critical_curves}
\end{figure}

\begin{figure}
    \centering
    \includegraphics[width=\linewidth]{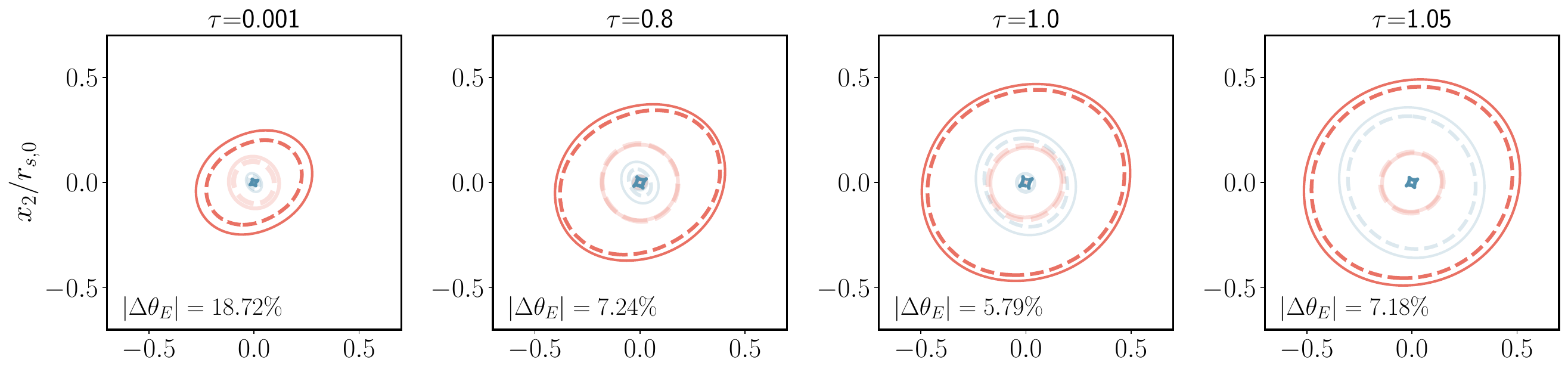}
    \includegraphics[width=\linewidth]{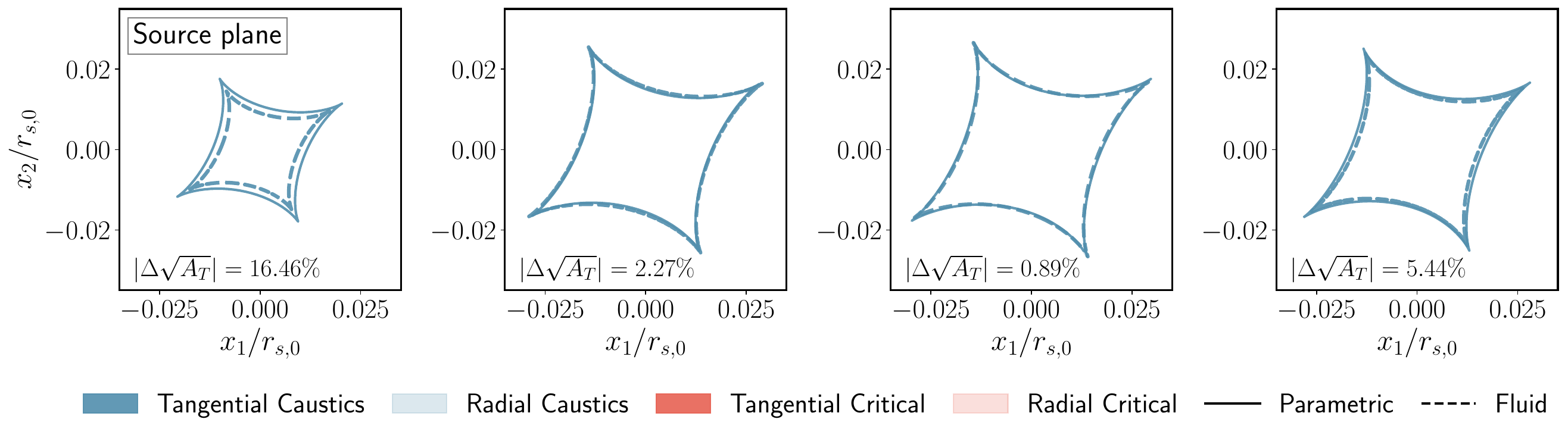}
    \caption{ Critical curves and caustics for an isolated SIDM halo with a constant ellipticity $e=0.6$ (or $b/a=0.8$). From left to right, the results correspond to progressively larger gravothermal phases with $\tau = 0.001$, $0.8$, $1.0$, and $1.05$. 
    Solid curves denote predictions from the parametric model, while dashed curves represent numerical results from the fluid simulation. Tangential critical curves and caustics are highlighted with darker lines, while radial critical curves and caustics are shown with lighter lines for clarity. 
    The lower panel provides a close-up illustration of the tangential caustics. $|\Delta \theta_E|$ is defined consistently with Figure~\ref{fig:critical_curves} and $A_T$ refers to the area enclosed by the tangential caustic.
    }
    \label{fig:critical_curves_ell}
\end{figure}

Figure~\ref{fig:critical_curves} compares the model-predicted and numerically simulated critical curves (red) and caustics (blue) for a spherically symmetric halo at $z_l = 0.439$ with a source plane at $z_s = 3$. 
Given our setup, the surface densities within the range $0.019 \lesssim \tau \lesssim 0.45$ remain below $\Sigma_{\rm cr}$ for $r > 0.1 r_{s,0}$. 
Therefore, we present results for $\tau = 0.001$, $0.8$, $1.0$, and $1.05$, from left to right.
Our findings show excellent agreement between the two predictions. The effective Einstein radius is defined as $\theta_E = \sqrt{A_E / \pi}$, where $A_E$ represents the area enclosed by the tangential critical curve.
A discrepancy of 7\% in the Einstein radius is observed only at $\tau = 0.8$. Due to the assumed spherical symmetry, the tangential caustics in the source plane degenerate into single points at the origin, rather than forming extended curves. These correspond to circular critical curves in the image plane.

Importantly, non-zero ellipticity plays a critical role in lensing studies, as it introduces
non-vanishing tangential caustics, enabling the formation of four-image lensing events. In this context, galaxy-galaxy strong lensing (GGSL) cross section quantifies the area on the source plane within which a background source would be multiply imaged due to the lensing potential of a foreground halo. It is highly sensitive to both the inner mass distribution and ellipticity of the lens. In SIDM halos, ellipticity is influenced by the thermalization of the inner regions, which reduces ellipticity compared to the CDM case in a radius-dependent manner. 

Exploring the effects of a radius-dependent ellipticity is beyond the scope of this work. Instead, we introduce a constant ellipticity using the coordinate transformation method of Ref.~\cite{r.kormann*p.schneiderandm.bartelmannIsothermalEllipticalGravitational1993}, and numerically compute the resulting lensing predictions. The ellipticity is defined as $e = \sqrt{1 - q^2}$, where $q = b/a$ is the axis ratio of the projected mass distribution.
Figure~\ref{fig:critical_curves_ell} presents the results for an ellipticity of $e=0.6$ (or $q=b/a=0.8$). Compared with the spherical case, we observe the emergence of diamond-shaped tangential caustics in the inner regions. Aside from this, the other curves are distorted in shape while maintaining approximately the same area. To better visualize the tangential caustics, we provide zoomed-in views of them in the bottom panels.
Comparing the panels at different $\tau$ values, we find that core-collapsing halos, represented by the last three panels, exhibit larger tangential GGSL cross sections than the leftmost panel, which closely resembles the NFW case.

\begin{figure}
    \centering
    \includegraphics[width=0.465\linewidth]{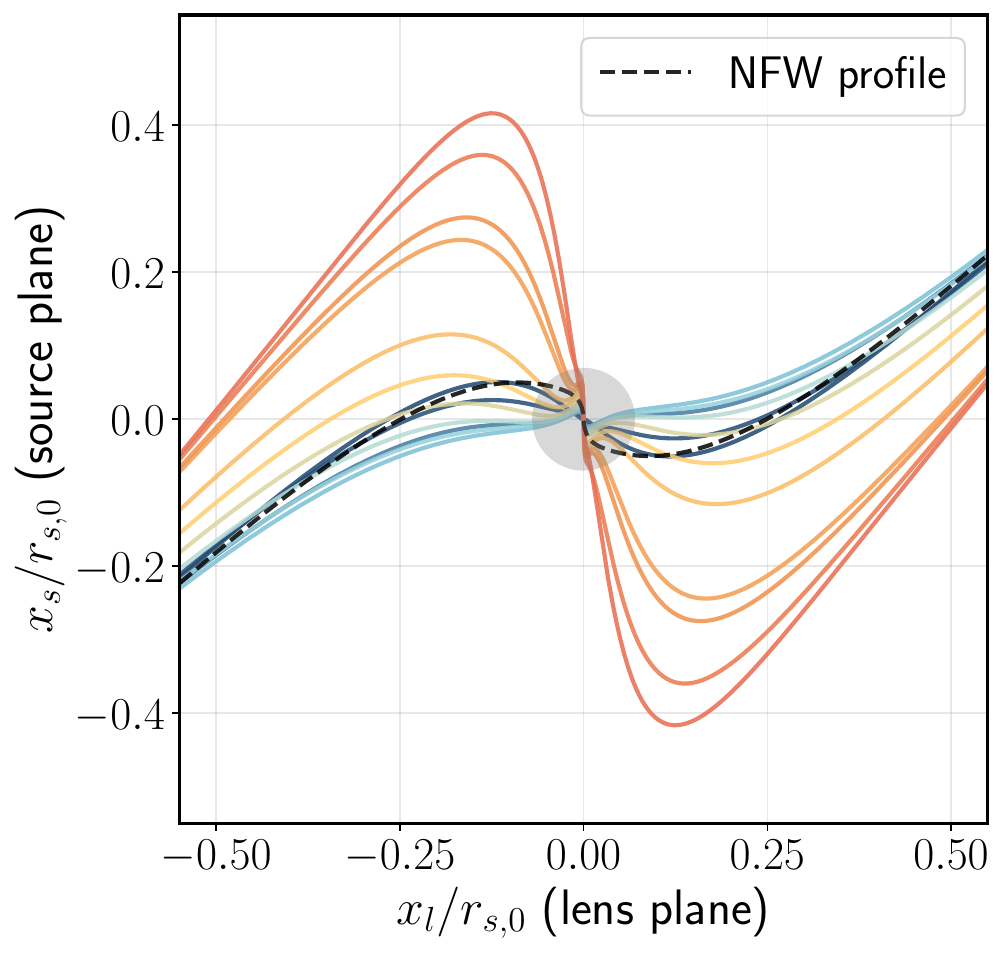}
    \includegraphics[width=0.525\linewidth]{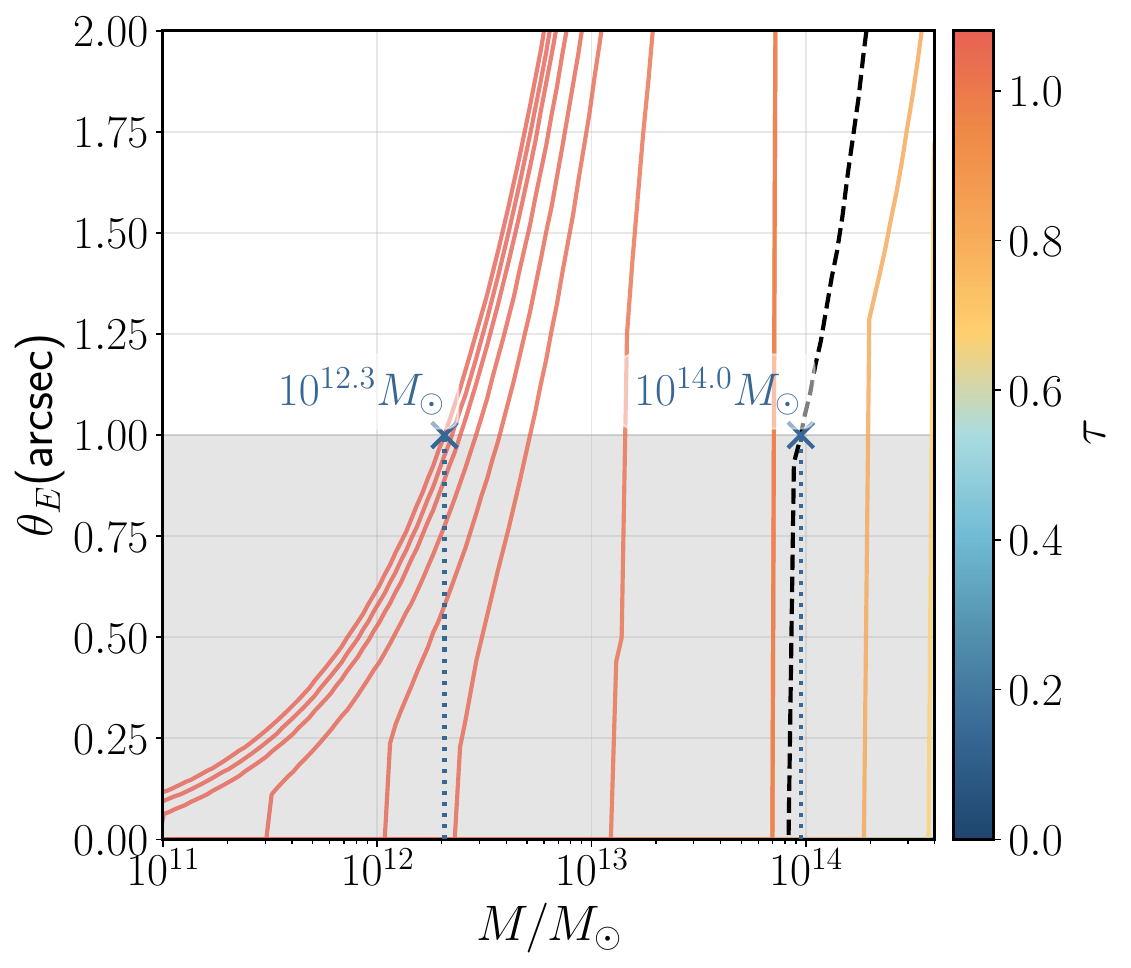}
    \caption{
        Lensing features of isolated spherical SIDM halos as functions of gravothermal phase and mass.
        \textbf{Left panel:} The source plane coordinate $\hat{x}_s = x_s/r_{s,0}$ as a function of the image plane coordinate $\hat{x}_l = x_l/r_{s,0}$, derived from the lensing equation. The NFW case at $\tau = 0$ is shown as a dashed black curve, while the curves for finite $\tau$ values are color-coded according to the bar on the right. The shaded gray region highlights the parameter space where the parametric lensing model's relative accuracy may exceed 20\%. \\
        \textbf{Right panel:} The Einstein radius ($\theta_E$) as a function of the halo mass ($M$) at different $\tau$ values as color-coded in the bar. 
        The region with $\theta_E<1~$arcsec is shaded in gray, suggesting no detectable signatures. 
        We mark the minimum detectable halo mass in the NFW case ($\tau=0$) and the core collapsed case ($\tau=1.08$) with labeled crosses, revealing that SIDM can reduce this threshold by more than one order of magnitude, from $10^{14.0} M_\odot$ to $10^{12.3} M_\odot$.
    }
    \label{fig:density_lensing}
\end{figure}

To further illustrate the impact of SIDM on the lensing signatures of isolated halos, we investigate how light rays are bent under a spherically symmetric SIDM lens and how this influences the statistics of lensing events.

In the left panel of figure~\ref{fig:density_lensing}, we plot the lensing equation at various $\tau$ values, with colors corresponding to the bar on the right. The lensing equation,
\begin{equation}
x_s = x_l - \frac{D_{LS}}{D_S} \alpha_i,
\end{equation}
relates the source plane coordinates $\mathbf{\hat{x}}_s = \mathbf{x}_s/r_{s,0}$ to the lens plane coordinates $\mathbf{\hat{x}}_l = \mathbf{x}_l/r_{s,0}$, encapsulating information about the deflection angle $\mathbf{\alpha}$. Starting with an NFW halo (dashed black) corresponding to $\tau=0$, the SIDM-induced gravothermal evolution reduces the lensing effect in the early stages and then enhances it.  
During the core formation phase, depicted in blue hues, the Einstein radius---identified as the intersection with the x-axis ($x_s=0$)---initially decreases and can reduce to zero. 
As the core begins to shrink, the Einstein radius grows once more, eventually exceeding the original NFW value at $\tau \gtrsim 0.6$, as highlighted by the warmer colors.

In the right panel, we plot the Einstein radius as a function of halo mass, revealing correlations at different $\tau$ values. Introducing a minimum resolvable Einstein radius of one arcsecond, we show that a core-collapsed SIDM halo can lower the minimum detectable mass from $10^{14.0} M_\odot$ in CDM to $10^{12.3} M_\odot$ in SIDM. This significant shift in the detection threshold could profoundly impact the statistics of lensing events, as lower-mass halos are more abundant due to the nearly power-law nature of the halo mass function (approximately $M^{-2}$).
In practice, the fraction of core-collapsed halos depends on the SIDM model and is influenced by the baryon content. The latter aspect will be explored further in Section 5.

\subsection{Subhalos in a host halo}
\label{sec:SIDM Dynamics of Cluster Halo with Embedded Central Subhalo}

Subhalos embedded within a host halo experience external shear from the host halo, leading to finite tangential caustics even when the subhalo is modeled as spherically symmetric. This shear contribution, dominated by the host halo's deep gravitational potential, significantly influences the lensing features of the subhalo. Notably, without the host halo's contribution, the subhalo's potential alone is insufficient to generate a large Einstein radius capable of producing detectable lensing signatures. Thus, the host halo's influence is critical in shaping the observable lensing properties of subhalos.

\begin{table}[h]
    \centering
    \renewcommand{\arraystretch}{1.2} 
    \setlength{\tabcolsep}{10pt} 
        \begin{tabular}{|l|r|r|}
        \hline
        \textbf{Parameter} & \textbf{Main Halo} & \textbf{Subhalo} \\
        \hline
        Mass ($\rm M_\odot/h$) & $5 \times 10^{15}$ & $1 \times 10^{13}$ \\
        Scale Radius, $r_{s,0}$ (kpc/h) & 933 & 75 \\
        Density, $\rho_{s,0}$ ($\rm h^2 M_\odot$/kpc$^3$) & $8.92 \times 10^{5}$ & $2.32 \times 10^{6}$ \\
        \hline
    \end{tabular}
    \caption{Parameters for the main halo and subhalo at $\tau=0$. The redshifts for the lens and source are $z_{\text{lens}} = 0.439$ and $z_{\text{source}} = 3$, respectively. }
    \label{tab:halo_parameters}
\end{table}

In this section, we evaluate the effectiveness of a parametric lensing model in describing the critical curves and caustics of subscale lenses. We focus on velocity-dependent SIDM models, which decouple the effective SIDM cross-sections of the host halo and the subhalo. This decoupling allows the host halo to remain unaffected by SIDM effects while enabling the subhalos to evolve into deep core-collapse phases. 
For clarity, we assume the host halo follows an NFW profile (i.e., $\tau = 0$), and we explore the lensing features of subhalos at representative gravothermal evolution phases and varying distances from the host halo. 
Table~\ref{tab:halo_parameters} summarizes the parameters of the main halo and subhalo used in this study, chosen to enhance strong lensing features and more clearly demonstrate the accuracy of our calibrated model. The main halo adopts the same NFW parameters as the isolated halo discussed in the previous section. The lens redshift ($z_l = 0.439$) and source redshift ($z_s = 3$) are also kept consistent with the previous setup. To obtain the Einstein radius $\theta_E$ of the host halo, we first identify the radius of the tangential critical curve from the results in figure~\ref{fig:critical_curves}, and use this radius as the Einstein radius of the main halo. 
\begin{figure*}
    \centering
    \includegraphics[width=\linewidth]{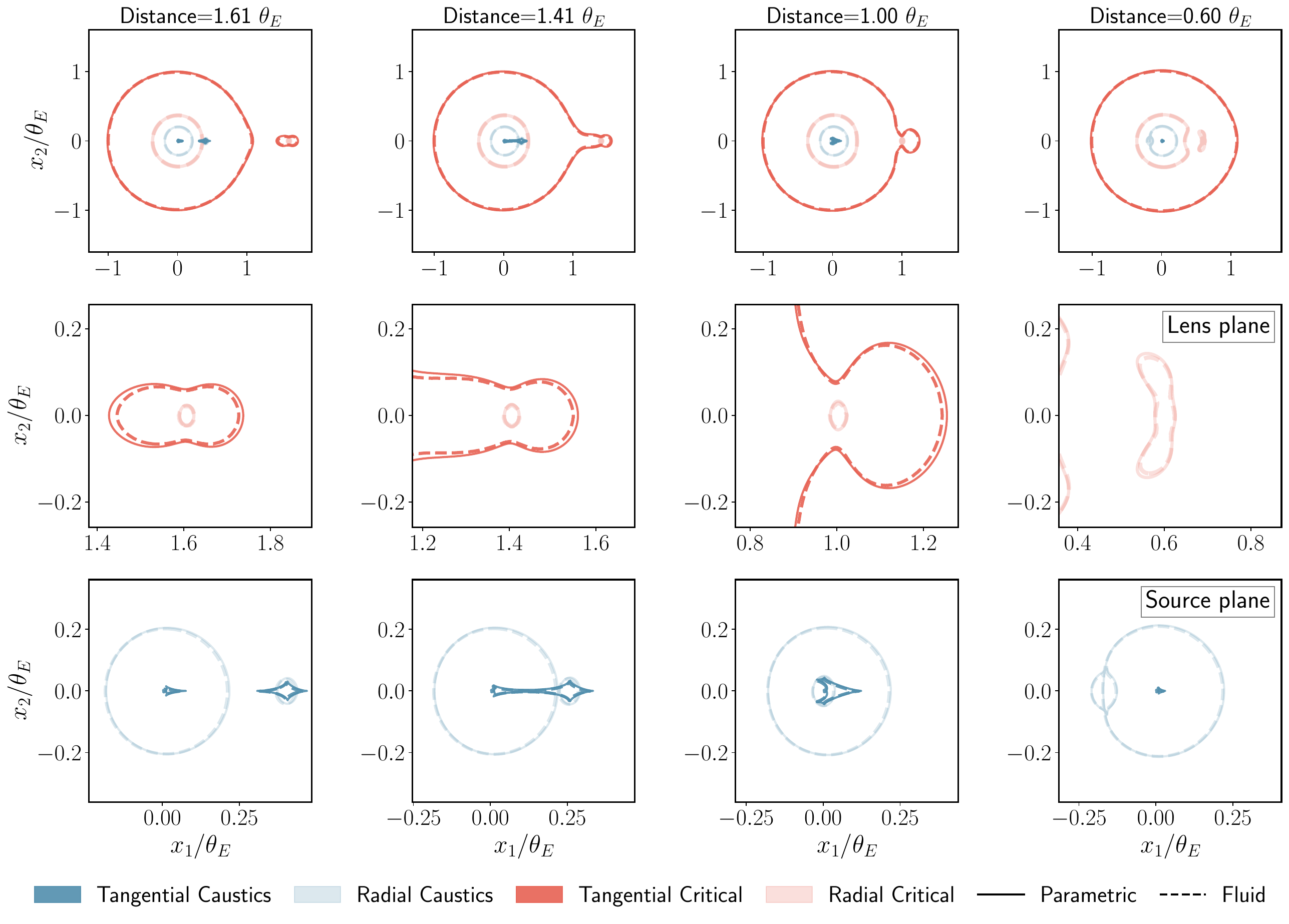}
    \caption{Lensing features of a subhalo with $\tau = 1.08$ embedded in a host halo with $\tau = 0$, shown at separations of $1.61$, $1.41$, $1.00$, and $0.60$, measured in units of the main halo's Einstein radius. The top row presents an overview of both systems, with critical curves and caustics depicted in red and blue, respectively. The middle and bottom rows provide zoomed-in views, enhancing the visualization of substructures. Darker lines indicate tangential critical curves and caustics, while lighter lines represent radial ones. Across all panels, the solid lines represent predictions from the parametric model, while dashed lines depict results from fluid simulations, illustrating the degree of agreement between the two approaches.
    }
    \label{fig:mainhalo_critical}
\end{figure*}
Figure~\ref{fig:mainhalo_critical} (top panel) illustrates how the tangential (opaque) and radial (light) critical curves (red) and caustics (blue) evolve as the distance between the subhalo and host halo changes, with distances expressed in terms of the host halo's Einstein radius. The middle and bottom panels provide zoomed-in views of the critical curves and caustics on the lens and source planes, respectively.
When the subhalo is far from the host halo (left panels), its critical curves and caustics remain distinct and separate from those of the host halo. In this configuration, the host and subhalo potentials induce external shear on each other, resulting in non-zero areas enclosed by the tangential caustics (tangential GGSL cross section), which is particularly prominent for the subhalo.
As the subhalo approaches the host halo’s Einstein radius, the total surface density more easily exceeds the critical surface density $\Sigma_{\rm cr}$, leading to more pronounced distortions in both types of curves. However, as the subhalo moves even closer to the host halo, its critical curves and caustics begin to merge with those of the host, and the subhalo contributes to perturbations in the lensing signatures originally induced by the host halo.
In all cases, we present numerical results from the fluid simulation as dashed curves, demonstrating their good agreement with our model predictions.

\begin{figure*}
    \centering
    \includegraphics[width=\linewidth]{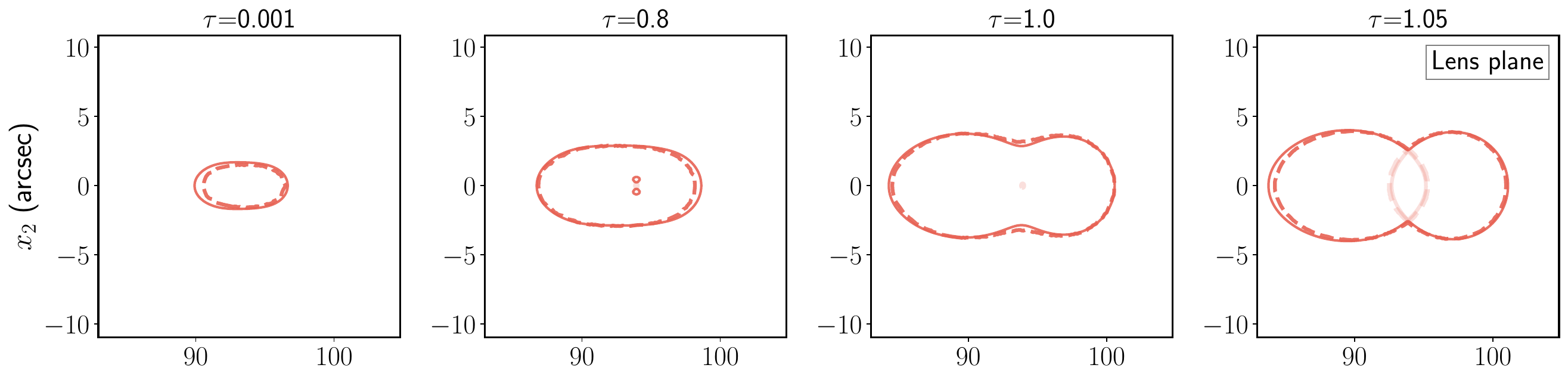}
    \includegraphics[width=\linewidth]{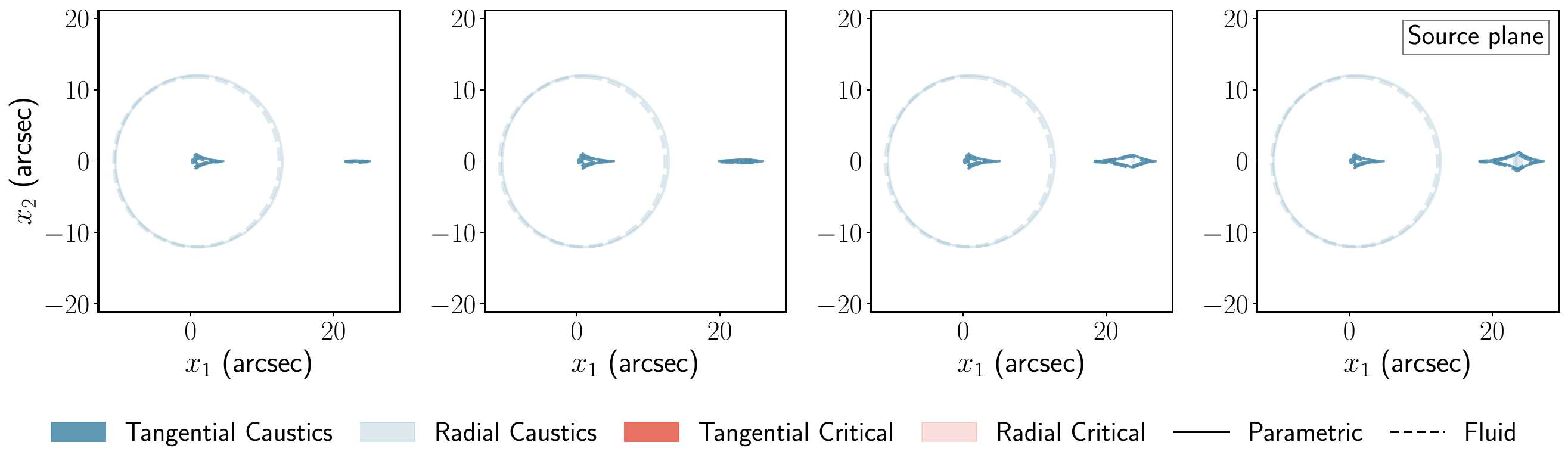}
    \caption{
Lensing features of subhalos at different gravothermal phases within an NFW host halo, located at a separation of $1.61$ times the main halo's Einstein radius. The top row shows zoomed-in critical curves on the lens plane for subhalos with $\tau = 0.001$, $0.8$, $1.0$, and $1.05$. The bottom row displays the corresponding caustics on the source plane. Darker lines represent tangential critical curves and caustics, while lighter lines indicate radial ones. Solid lines show predictions from the parametric model, and dashed lines represent results from fluid simulations, highlighting the agreement between the two methods.
    }
     \label{fig:subhalo_critical_1.61}
\end{figure*}
\begin{figure*}
    \centering
    \includegraphics[width=\linewidth]{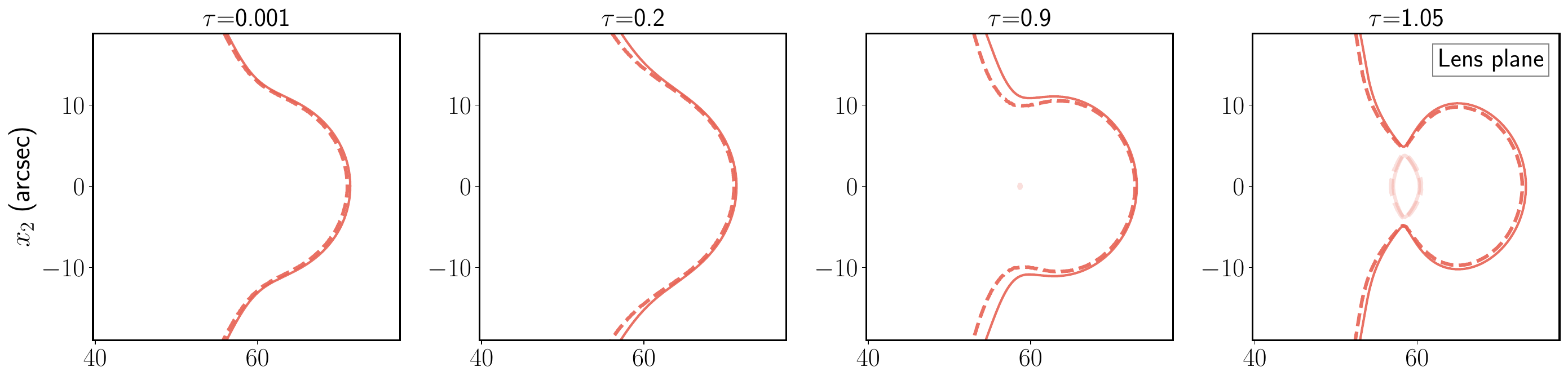}
    \includegraphics[width=\linewidth]{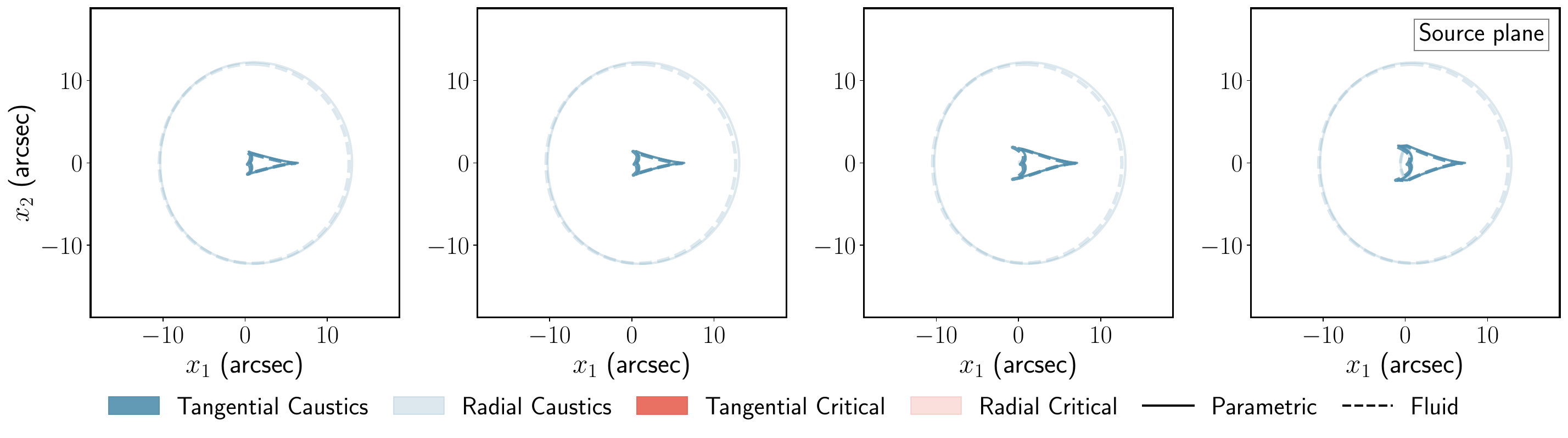}
    \caption{As in figure~\ref{fig:subhalo_critical_1.61}, this figure shows the subhalo residing at the main halo's Einstein radius, with separate panels illustrating the subhalo at $\tau=0$, $0.2$, $0.9$, and $1.05$.
    }
     \label{fig:subhalo_critical_1}
\end{figure*}

In figures~\ref{fig:subhalo_critical_1.61} and~\ref{fig:subhalo_critical_1}, we examine how the tangential (opaque) and radial (light) critical curves (red) and caustics (blue) evolve at different values of $\tau$. 
Figure~\ref{fig:subhalo_critical_1.61} shows a subhalo located at $1.61 \theta_{E}$ from the host halo, a separation large enough to distinguish the curves associated with the subhalo and the host. For cases that are not deeply core-collapsed, there could be no surface densities exceeding the $\Sigma_{\rm cr}$, so we focus on core collapsed cases at $\tau = 0.8$, $1.0$, and $1.05$, in addition to the left panel, which is close to the initial condition. At $\tau \geq 0.8$, we observe notable changes in the lensing signatures, including the formation of a dumbbell-shaped critical curve at $\tau \gtrsim 1.0$, and an increasing tangential GGSL cross section as $\tau$ increases.

Figure~\ref{fig:subhalo_critical_1} depicts a subhalo located at the Einstein radius of the host halo. This separation is too small to fully separate the lensing signatures of the two halos. However, the larger total surface density allows for an exploration of lensing signatures at smaller values of $\tau$, so we consider $\tau = 0.001$, $0.2$, $0.9$, and $1.05$. From the critical curves (top panels), we note that larger $\tau$ values cause the curves to bend more significantly, making the lensing signatures easier to identify. In the caustics (bottom panels), the smallest tangential GGSL cross section (for the combined host and subhalo) occurs at $\tau = 0.2$, while it grows considerably at $\tau = 0.9$ and $1.05$. 
In both figures, the results from the parametric model and the fluid model align very well.

\section{Lensing signatures of core collapsed halos in the self-similar regime}
\label{sec:selfsimilar}

In the conducting fluid model of SIDM halos, self-interactions are modeled through a heat transport equation, where the scattering cross section determines the thermal conductivity $\kappa$. Unlike ordinary fluids, SIDM halos primarily exhibit a long-mean-free-path (LMFP) regime, in which particles can orbit the halo center multiple times after a collision, suppressing the energy transport. This contrasts with the short-mean-free-path (SMFP) regime typical of fluid dynamics. To capture both regimes and their transition during gravothermal evolution, the effective conductivity is defined as $\kappa_{\rm eff} = \left( \kappa_{\rm LMFP}^{-1} + \kappa_{\rm SMFP}^{-1} \right)^{-1}$, providing a smooth interpolation between the LMFP and SMFP regimes~\cite{balberg:2001qg,koda11013097,pollack:2014rja,balberg:2002ue,essig:2018pzq,nishikawa:2019lsc,yang220503392}.

If an SIDM halo has evolved for a time exceeding its expected core collapse time, its configuration extends beyond the scope of our current modeling, necessitating dedicated investigation. This is particularly relevant for models with large SIDM cross-sections, where the population of deeply core-collapsed halos can be substantial, significantly influencing our expectations of lensing signatures in observations.

In the literature, a halo evolving deeply into the core collapse regime, corresponding to $\tau\gtrsim 1$ in our parametric framework, is demonstrated to depict a self-similar solution, where the inner core contracts with the extended outer halo remains almost static, maintaining a power-law density profile described by $r^{-u}$ with $u$ ranges between $2$ and $2.5$. Numerical simulations indicate a preferred value of $u\approx 2.2$~\cite{Lynden-Bell:1980xip,balberg:2001qg}.

\begin{figure}
    \centering
    \includegraphics[width=\linewidth]{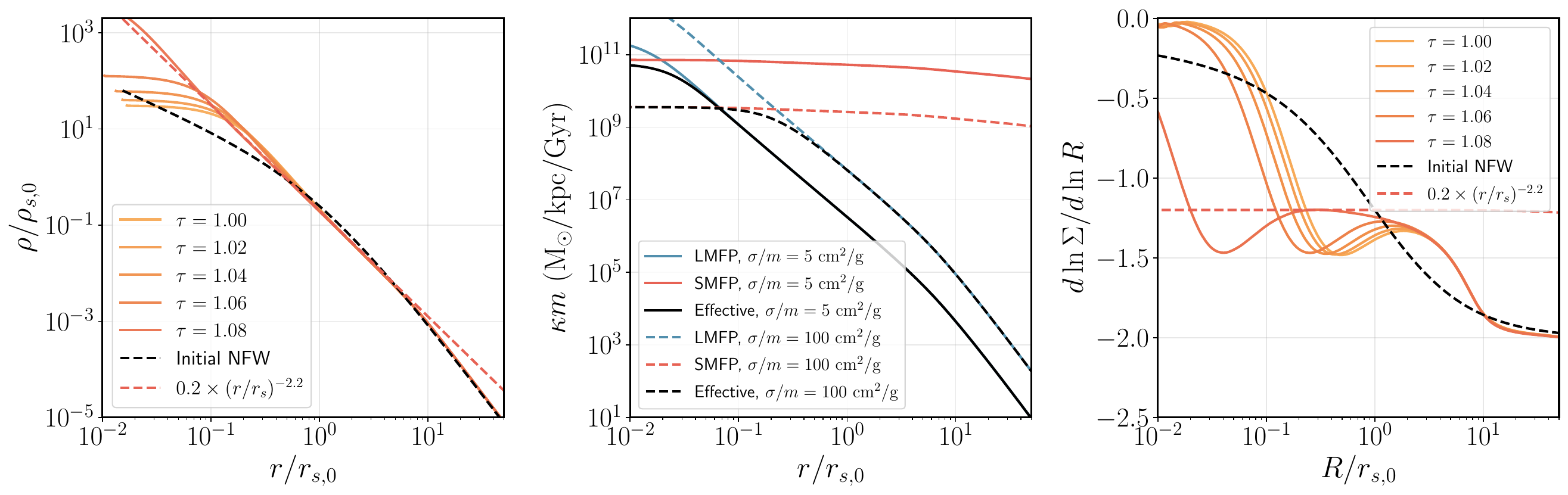}

    \caption{\label{fig:selfsimilar} Evolution of core-collapsing halo density profiles in the self-similar regime, corresponding to $\tau\gtrsim 1$. The left panel shows that, apart from the inner cores and the outer regions with $r\gg r_s$, the density profiles remain nearly static and close to a power-law form, $r^{-2.2}$ (dashed blue). For reference, the initial NFW profile (dashed black) is also plotted. The middle panel highlights the distinction between the LMFP and SMFP regimes. The effective heat conductivity (black) approximately follows the lower value of the LMFP (blue) and SMFP (red) conductivities. For $\sigma/m=5~\rm cm^2/g$ (solid) and $\sigma/m=100~\rm cm^2/g$ (dashed), the LMFP regime corresponds to $r/r_s>0.02$ and $r/r_s>0.2$, respectively. The right panel presents the logarithmic slopes of the surface densities for the profiles in the left panel, illustrating that the inner cores—characterized by the minima at $R/r_s<1$---contract progressively faster with increasing $\tau$ while maintaining a self-similar shape in the region $r\lesssim r_s$.}  
\end{figure}

In the left panel of figure~\ref{fig:selfsimilar}, we show the evolution of halo density profiles from $\tau = 1$ to $1.08$ and compare them to a power-law distribution of $r^{-2.2}$. As the core contracts, the inner halo density increases, while the region between the core and the outer NFW-like halo remains close to the $r^{-2.2}$ profile.

The middle panel presents the distribution of $\kappa m$ for the effective (black), LMFP (blue), and SMFP (red) scenarios, for the final snapshot at $\tau = 1.08$. This plot demonstrates that the outer halo remains in the LMFP regime even during the late stages of core collapse. Notably, the transition between the two regimes depends on the SIDM model. For the $\sigma/m = 5~\rm cm^2/g$ case (solid lines), which we adopted in the fluid simulation, the transition occurs at $r/r_s \approx 0.02$. Assuming universality and hydrostatic equilibrium, we can extend this result to other SIDM models by holding the simulated halo properties fixed. For instance, for $\sigma/m = 100~\rm cm^2/g$ (dashed lines), the transition shifts to $r/r_s \approx 0.2$, an order of magnitude larger.

In the right panel, we plot the logarithmic slopes of the surface densities for the profiles in the left panel. This reveals that the inner cores, characterized by minima at $R/r_s < 1$, contract progressively faster with increasing $\tau$, while maintaining a self-similar shape in the region $r \lesssim r_s$.
Compared with the initial NFW profile, the self-similar evolution results in systematically steeper logarithmic slopes in core-collapsed halos.

Based on existing studies, we assume that halos in the LMFP regime stop evolving at $t \gtrsim 1.1 t_c$ within our parametric framework. In contrast, the inner halo regions in the SMFP regime do not evolve universally across different SIDM models and halos, introducing an irreducible systematic uncertainty. Nevertheless, as shown in the middle panel of figure~\ref{fig:selfsimilar}, the SMFP region remains deep within $r_s$, exerting only a limited impact on observable features.

\begin{figure}
    \centering
    \includegraphics[width=\linewidth]{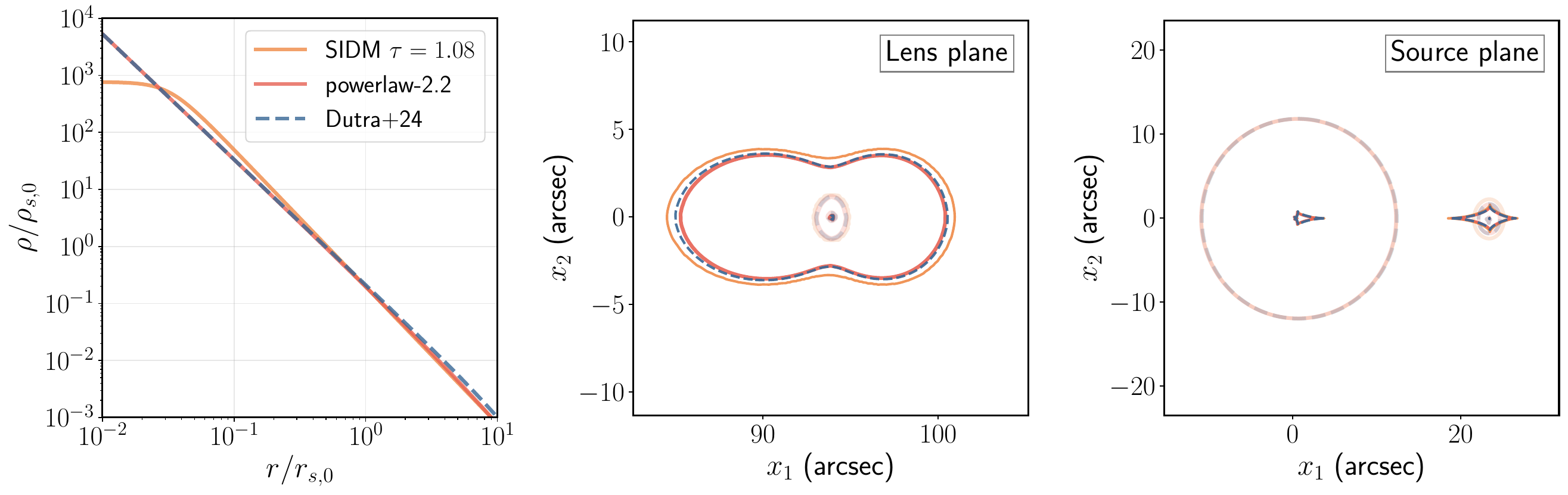}
    \caption{\label{fig:diffSMFP} Comparison of lensing features in different models of core-collapsed halos: the parametric model at $\tau=1.08$ (orange), power-law-2.2 (red), and Dutra+24~\cite{Dutra:2024qac} (blue). The left panel shows their density profiles, while the middle and right panels depict the critical curves and caustics, with darker lines representing tangential quantities and lighter lines indicating radial ones.}
\end{figure}

To assess how different descriptions of core-collapsed halos affect lensing features, we compare predictions from the simulated $\tau = 1.08$ case with two idealized power-law models. 
The first model, labeled \textit{power-law-2.2}, adopts an inner density profile proportional to $r^{-2.2}$, smoothly matched to the outer region of $\rho_{\rm PSIDM}$. 
The second model, following Dutra et al.~\cite{Dutra:2024qac}, uses a generalized NFW profile~\cite{Munoz:2001bw}:
\begin{eqnarray}
\rho_{\rm gNFW} = \frac{\rho_0}{(r/r_0)^{\gamma_1} \left(1 + (r/r_0)^2\right)^{(\gamma_2 - \gamma_1)/2}}, 
\end{eqnarray}
where we set $\gamma_1 = 2.2$ and $\gamma_2 = 3$, allowing $r_0$ and $\rho_0$ to vary in the fit. Figure~\ref{fig:diffSMFP} compares the density profiles (left), critical curves (middle), and caustics (right) for the fluid simulation at $\tau = 1.08$ (orange), \textit{power-law-2.2} (red), and \textit{Dutra+24} (blue).
We find that the $\tau = 1.08$ case introduces only minor differences at the percent level in the radial positions of both the critical curves and caustics, while the two power-law models produce nearly identical results.

\section{Impact of baryons}
\label{sec:baryons}

Thus far, our parametric model has been applied exclusively to dark matter halos, neglecting the influence of baryons. However, baryonic matter can significantly impact lensing signatures~\cite{yang:2021kdf,Yang:2024tba}. 
Due to its dissipative nature, baryonic matter tends to concentrate in the dense central regions of halos, forming compact galaxies. 
Since gravitational lensing is most sensitive to high-density regions where the surface density exceeds the critical threshold, the baryonic fraction enclosed within these regions can be substantially higher than its cosmological average. 
In some galaxies, the baryonic mass enclosed within the Einstein radius may be comparable to or even exceed that of dark matter.

Aside from modifying the mass distribution, baryons also affect the shape of density profiles. Their presence induces galaxy-dependent contraction effects, altering the profile from its dark-matter-only form and complicating the interpretation of lensing signals in realistic galaxy-halo systems. Moreover, in SIDM scenarios, several studies have shown that baryons can accelerate the gravothermal evolution of halos. An increased fraction of core-collapsed halos can enhance observable lensing signatures, potentially addressing the debated excess of small-scale lenses in observations.

In this section, we investigate the influence of baryons on lensing-related features. We adapt the approach introduced in ref.~\cite{Yang:2024tba} for the contracted-$\beta4$ profile to the PSIDM profile, incorporating both the boost in gravothermal evolution and the contraction of core size. The acceleration of the evolution is implemented through the ratio of core-collapse timescales with and without baryons:
\begin{eqnarray}
{\cal F}_t\equiv t_{cb}/t_c = \left( \frac{1}{\hat{r}_{\rm eff}} + \frac{ 20 \hat{\rho}_H \hat{r}_H^3}{\hat{r}_{\rm eff}(\hat{r}_{\rm eff}+\hat{r}_H)^2} \right)^{-1} \left( 1+\alpha\frac{\hat{\rho}_H \hat{r}_H^2}{2} \right)^{-\frac{1}{2}},
\end{eqnarray}
where hatted quantities are normalized by the NFW scale density $\rho_s$ and radius $r_s$. The baryonic content is modeled using a Hernquist profile with scale density $\rho_H$ and radius $r_H$. Using these parameters, we compute
$\hat{r}_{\rm eff} = (1+1.6 \hat{\rho}_H \hat{r}_H^3/2)/(1+1.6 \hat{\rho}_H \hat{r}_H^2/2)$.

To construct the contracted-PSIDM profile in analogy to the contracted-$\beta4$ profile, we use $t_{cb}$ to determine $\tau$ and shrink $r_c(\tau)$ by a factor of ${\cal F}_t^2$. Our approach neglects baryon-induced distortions in the density profile, which could be incorporated using the Cored-DZ profile from ref.~\cite{Yang:2024tba}. However, the simplified method adopted here effectively captures the major baryonic effects and is sufficient to illustrate the key points of this analysis.

\begin{figure}
    \centering
    \includegraphics[width=0.47\linewidth]{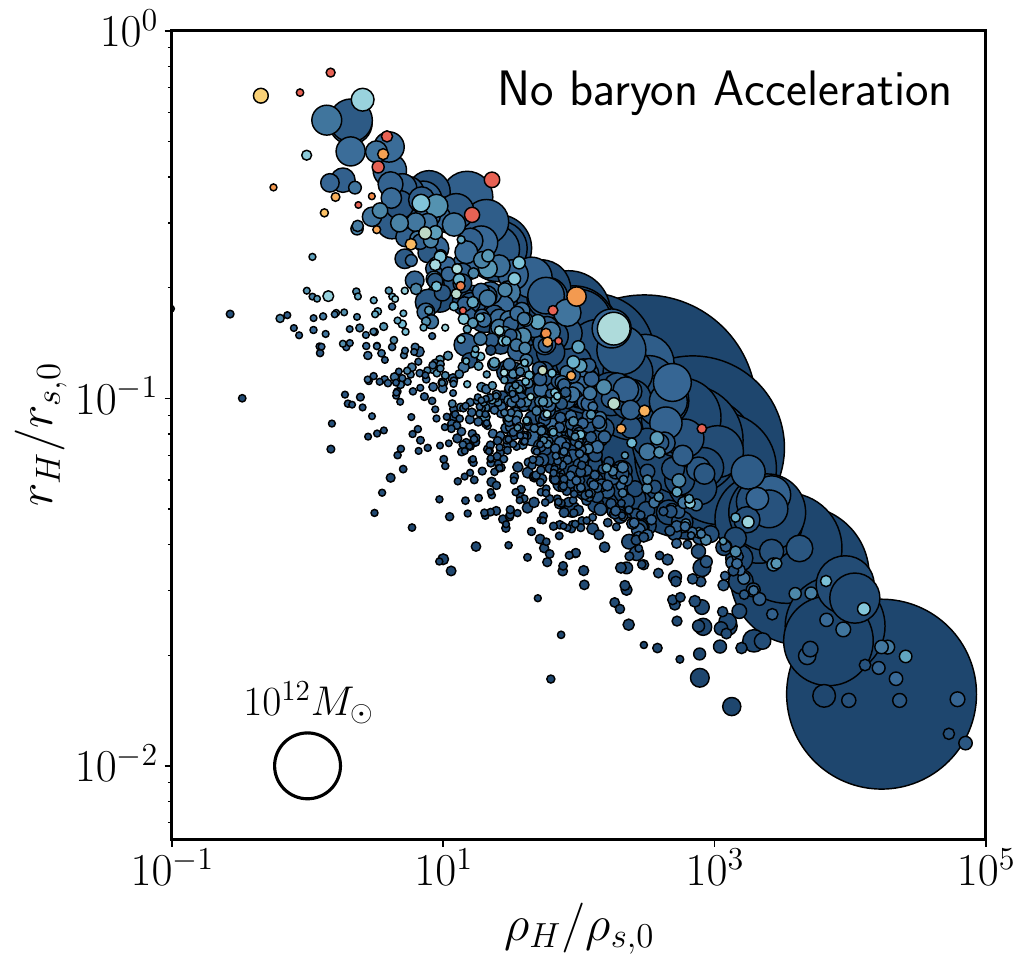}
    \includegraphics[width=0.52\linewidth]{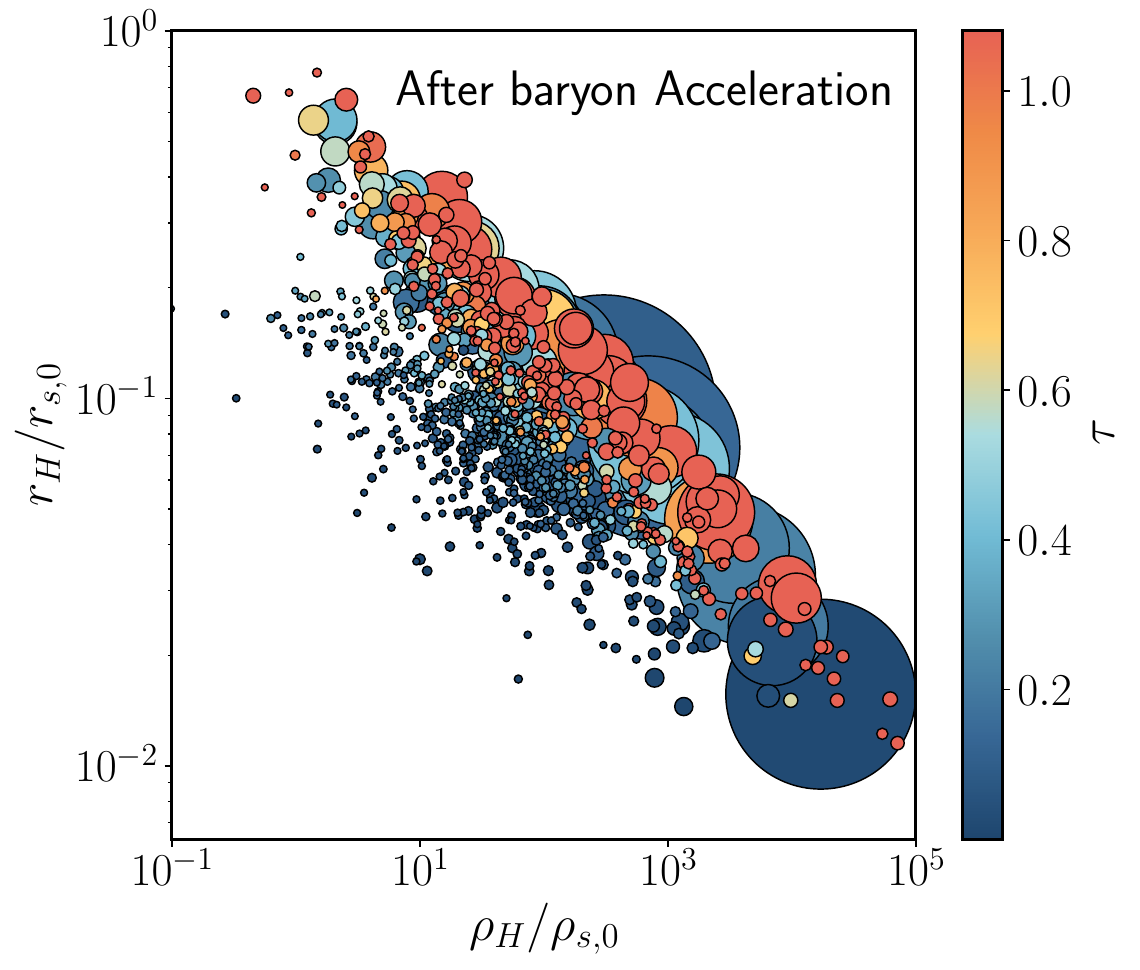}
    \caption{The distribution of gravothermal phases, represented by color-coded circles with sizes proportional to halo masses, on the size-density plane ($\rho_H/\rho_s - r_H/r_s$) of baryons relative to dark matter. The SIDM model considered here assumes $\sigma_0/m = 100~\rm cm^2/g$ and $w = 60~\rm km/s$, parameterizing a Rutherford-like scattering cross section. The left and right panels show results for halos excluding and including baryon effects, respectively, highlighting the significant boost in the gravothermal phase due to the presence of baryons. The halo sample is randomly selected from isolated halos with masses exceeding $10^{10}\rm M_{\odot}$ in the TNG-50-1 simulation, with the most massive halo reaching approximately $10^{13}\rm M_{\odot}$. With baryonic acceleration included, the fraction of halos that reach the core-collapsed phase ($\tau \ge 1.0$) increases from 1.2\% to 21.2\%.
    }
    \label{fig:baryon_acceleration}
\end{figure}

\begin{figure}
    \centering
    \includegraphics[width=0.49\linewidth]{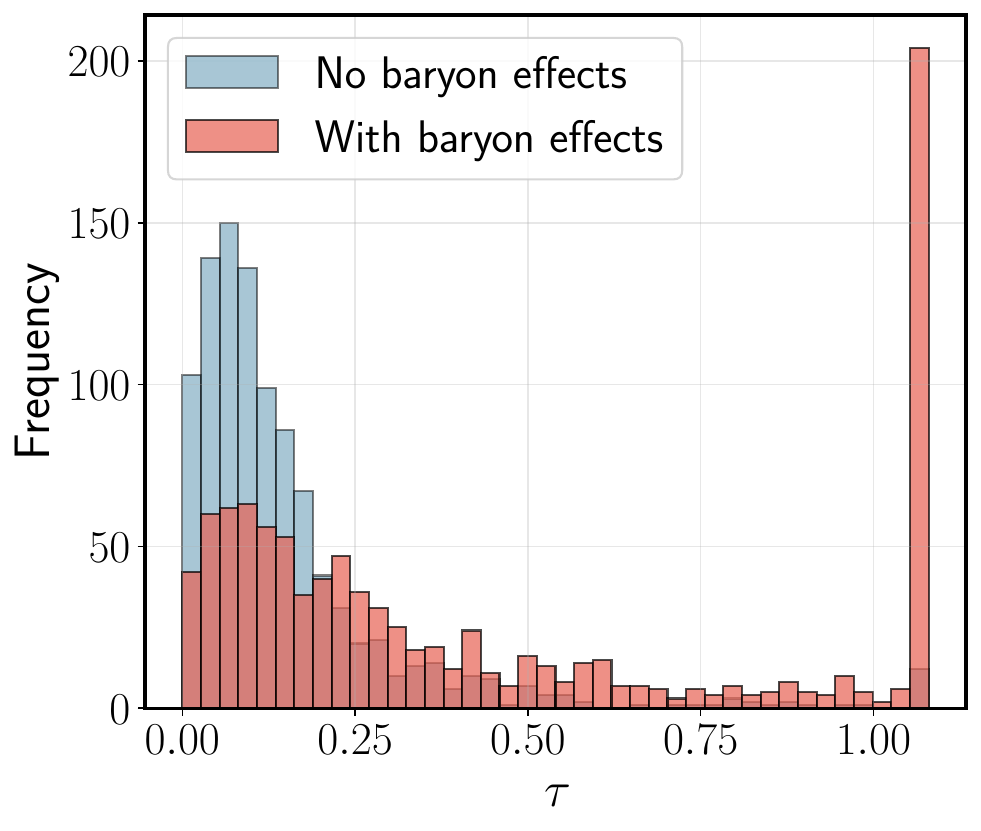}
    \includegraphics[width=0.49\linewidth]{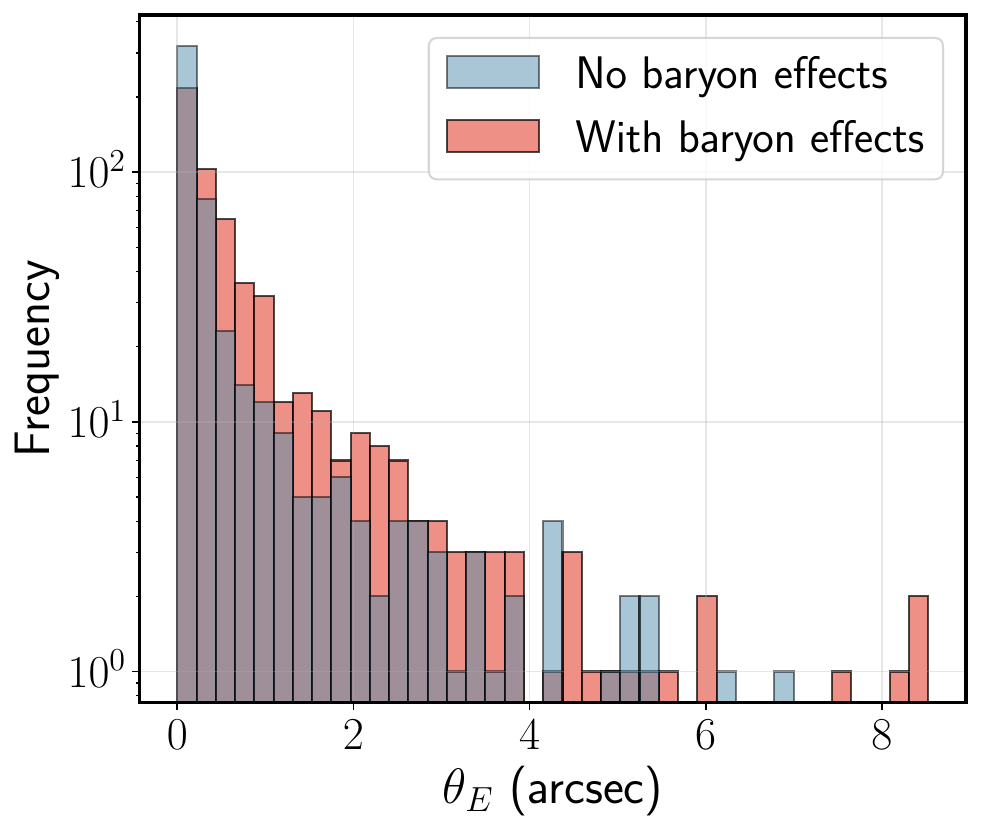}
    \caption{Distribution of $\tau$ (left) and Einstein radius $\theta_E$ (right) with (red) and without (blue) baryon effects for the halos in figure~\ref{fig:baryon_acceleration}. In the right panel, the fraction of halos with nonzero $\theta_E$ is 45\% with baryon acceleration and 49\% without it.}
     \label{fig:Einstein_radius_and_tau}
\end{figure}
In figure~\ref{fig:baryon_acceleration}, we compare the distribution of gravothermal phases with (right) and without (left) baryon acceleration. We randomly select 1000 isolated halos hosting galaxies with masses above $10^{10}\rm M_{\odot}$ from the TNG-50-1 simulation, modeling the halo and baryonic components using NFW and Hernquist profiles, respectively.
For the halo component, we determine the NFW scale parameters $\rho_s$ and $r_s$ by solving for the concentration using the virial radius and half-mass radius.
Similarly, for the baryonic component, we compute the Hernquist scale parameters $\rho_H$ and $r_H$ based on the total mass and half-mass radius.
Given the halo formation time $t_f$ (look-back time), we compute the gravothermal phase as $\tau = t_f/t_c$ in the left panel and $\tau = t_f/t_{cb}$ in the right panel, adopting a Rutherford-like SIDM model~\cite{feng09110422,Ibe:2009mk}:
\begin{eqnarray}
\label{eq:rxs}
\frac{d\sigma}{d \cos\theta} = \frac{\sigma_{0}w^4}{2\left[w^2+{v^{2}}\sin^2(\theta/2)\right]^2 },
\end{eqnarray}
where $\sigma_0/m = 100~\rm cm^2/g$ and $w = 60~\rm km/s$. Strictly speaking, this model applies to distinguishable initial states. See ref.~\cite{yang220503392} for a consistent particle interpretation in N-body simulations.
The results, plotted on the size-density plane ($\rho_H/\rho_s - r_H/r_s$), demonstrate that baryons can significantly enhance $\tau$ values, particularly for halos with a high baryon-to-halo mass ratio, which spread towards the upper-right region of the plane. Overall, the fraction of halos that reach the core-collapsed phase ($\tau \ge 1.0$) increases significantly due to baryonic effects, rising from 1.2\% to 21.2\%. This enhancement is most prominent in intermediate-mass halos and becomes less pronounced toward both lower-mass dwarf halos and higher-mass cluster halos---represented by smaller and larger blue circles, respectively---where the impact of baryons diminishes. In these mass regimes, our analytic lensing model remains more reliable. Quantitatively, within the mass range $10^{10} - 10^{11} \rm M_{\odot}$, the fraction of core-collapsed halos increases from 1.4\% in the absence of baryons to 19\% when baryonic effects are included. 
For higher mass bins, $(10^{11},10^{12})\rm M_{\odot}$ and $(10^{12},10^{13})~\rm M_{\odot}$, no core-collapsed halos are found without baryons, but the inclusion of baryons raises the fractions to 41\% and 5.0\%, respectively.

To investigate the impact of baryon-induced effects on lensing, we numerically compute the Einstein radii $\theta_E$ for the halos in figure~\ref{fig:baryon_acceleration}, assuming the same lens and source redshifts as in Section~\ref{sec:lensingfeatures}, i.e., $z_l = 0.439$ and $z_s = 3$.

The left panel of figure~\ref{fig:Einstein_radius_and_tau} presents a histogram of the $\tau$ values for the two cases shown in figure~\ref{fig:baryon_acceleration}, demonstrating that the inclusion of baryons significantly increases the fraction of core-collapsed halos under the considered SIDM model.
The right panel of figure~\ref{fig:Einstein_radius_and_tau} displays a histogram of Einstein radii for the same halo population. Notably, 49\% (45\%) of halos have zero Einstein radii. If we impose a resolution limit of 0.2 arcseconds on $\theta_E$, the fraction of halos exceeding this threshold is 40\% and 62\% in the DM-only and DM+baryon cases, respectively.

\section{Conclusion and discussion}
\label{sec:Conclusions}

In this work, we have developed a universal analytic model for SIDM halos, designed to facilitate lensing studies. By calibrating the gravothermal evolution of the normalized lensing potential with a simple functional form, we derive analytic expressions for deflection angles and surface densities across all gravothermal phases. 
Additionally, we introduce a new parametric density profile allowing improved modeling in the deep core-collapse phase. 
We have validated the lensing predictions from our model across all gravothermal phases, in both individual halos and cluster-scale halos with embedded subhalos. Furthermore, our model is computationally efficient and readily applicable to a large number of halos. For practical implementation, we provide example analysis code on the following GitHub repository: 

\href{https://github.com/HouSiyuan2001/SIDM_Lensing_Model}{https://github.com/HouSiyuan2001/SIDM\_Lensing\_Model}

To assess the accuracy of our model, we adopt a perspective that separates uncertainties into two categories: those arising from the parametrized hypersurface and those associated with the interfacing parameters. Our calibrated hypersurface itself is highly accurate and representative: for any SIDM halo composed solely of dark matter, we expect it to provide a reliable fit. 
On the other hand, uncertainty lies in the interfacing parameters---$\rho_s$, $r_s$, and $\tau$---can be much larger. Specifically, given an NFW halo with fixed $\rho_s$ and $r_s$, the determination of $\tau$ depends on estimating the core-collapse time (Eq.~\ref{eq:tc}), which introduces significant uncertainty. In the literature, this uncertainty is typically parameterized by a normalization factor $C$, which has approximately 10--30\% uncertainty. Accurately determining the gravothermal phase remains an open challenge warranting further study. 

This study focuses on halos in isolation. However, a simplified treatment of the integral approach in ref.~\cite{yang:2023jwn} allows incorporating accretion histories relatively easily. 
In the dark matter-only scenario, ref.~\cite{Yang:2024uqb} (Appendix C) demonstrates that the effects of accretion are predominantly encapsulated in the gravothermal phase $\tau$. Given a halo’s evolutionary history, e.g. parameterized by the NFW scale parameters $\rho_s(t)$ and $r_s(t)$, the gravothermal phase at time $t$ can be computed as
\begin{equation}
\tau(t) = \int_{t_f}^{t} \frac{d t'}{t_{c}[\sigma_{\rm eff}(t')/m,\rho_s(t'),r_s(t')]},
\end{equation}
where $\sigma_{\rm eff}(t')/m$ is an effective cross section determined by $\rho_s(t')$, $r_s(t')$, and an SIDM model~\cite{yang220503392,yang:2022zkd}. 
Substituting this $\tau(t)$ into our lensing model provides a reasonable leading-order prediction. However, one should bear in mind that its accuracy can be undermined in certain cases, particularly due to tidal mass loss in subhalos on highly radial orbits. In such scenarios, the SIDM halo mass may gradually deviate from its CDM counterpart, and even before significant mass loss occurs, differences in the outer density profile can emerge, potentially impacting the lensing predictions.

We have also demonstrated the impact of baryons on lensing through two key effects: accelerating gravothermal evolution and contracting density profiles. In general, the presence of baryons adds complexity to accurately modeling lensing signatures. The contraction of dark matter in response to a galaxy’s potential is not universal and can only be captured through semi-analytical approaches. Achieving more precise modeling, particularly for lensing applications, remains an ongoing challenge.

Another important factor influencing lensing predictions for SIDM halos is ellipticity. Gravitational lensing is sensitive to mass distributions, and even a small ellipticity can significantly enhance tangential lensing cross-sections.
Our analysis shows that the elliptical shape of an SIDM halo can produce notable observational signatures. However, our example remains highly conceptual, as it does not account for the radius-dependent nature of SIDM-induced ellipticity, where the inner regions tend to be more spherical. Future work is needed to fully understand these signatures.

The application of our model goes beyond SIDM. Any halo whose density profile can be fitted by the PSIDM model at a given $\tau$ (Eqs.~3.9, 3.10) can use our framework to obtain analytic lensing predictions. For example, Ref.~\cite{Yang:2025dgl} recently showed that two-component dark matter models with mass segregation can yield halo profiles well approximated by PSIDM. It is also possible to superimpose multiple PSIDM profiles to model more complex halo profiles. The lensing effects can then be obtained by summing the contributions from the individual components. 

Beyond theoretical applications, our model is well-suited for studying halos with minimal or negligible baryonic content. Observationally, dark halos can still leave detectable imprints through lensing, including flux ratio anomalies, distortions in the strong lensing images of host galaxies, and gravitational microlensing effects~\cite{Gilman:2021sdr,Vegetti:2009cz,Vegetti:2012mc,minor:2020hic,2011ApJ...729...49E,Croon:2020wpr,nadler:2023nrd}.Investigating these signatures in the context of SIDM could provide crucial insights into the interacting nature of dark matter. Continued efforts in this direction are essential to unveiling these hidden properties.

\section*{Acknowledgments}

D.Y. was supported in part by the National Key Research and Development Program of China (No. 2022YFF0503304), the Project for Young Scientists in Basic Research of the Chinese Academy of Sciences (No. YSBR-092). N.L. acknowledges the support of the science research grants from the China Manned Space Project (No. CMS-CSST-2021-A01), the Ministry of Science and Technology of China (No. 2020SKA0110100), and, the CAS Project for Young Scientists in Basic Research (No. YSBR-062). G.L. acknowledges the support from the Natural Science Foundation of China (NSFC) (No. 12333001). We also acknowledge the IllustrisTNG collaborations for making their simulation data publicly available.

\appendix

\bibliographystyle{unsrt}
\bibliography{References.bib}
\end{document}